\documentclass[aps,a4paper,floatfix,nofootinbib]{revtex4}

\usepackage[utf8]{inputenc}

\usepackage{graphicx,color}
\usepackage{amsmath,amssymb,amsfonts,amsthm,amscd,bm}
\usepackage{booktabs}
\usepackage{cancel}

\def\tilde{\widetilde}
\def\bar{\overline}
\def\hat{\widehat}
\def\*{\star}
\def\[{\left[}
\def\]{\right]}
\def\({\left(}      

\def\){\right)}

\def\zbar{{\bar{z} }}
\def\frac#1#2{\dfrac{#1}{#2}}
\def\inv#1{\dfrac{1}{#1}}
\def\half{\tfrac{1}{2}}
\def\d{\partial}

\def\2pi{\hbox{$2\pi i$}}

\def\dsl{\raise.15ex\hbox{/}\kern-.57em\partial}
\def\Dsl{\,\raise.15ex\hbox{/}\mkern-.13.5mu D}
\def\vep{\varepsilon}

      \def\CF{{\cal F}}

   \def\CN{{\cal N}}   \def\CO{{\cal O}}
      
\def\CS{{\cal S}}

\def\2pi{\hbox{$2\pi i$}}

\def\dsl{\raise.15ex\hbox{/}\kern-.57em\partial}
\def\Dsl{\,\raise.15ex\hbox{/}\mkern-.13.5mu D}
\font\numbers=cmss12
\font\upright=cmu10 scaled\magstep1
\def\stroke{\vrule height8pt width0.4pt depth-0.1pt}
\def\topfleck{\vrule height8pt width0.5pt depth-5.9pt}
\def\botfleck{\vrule height2pt width0.5pt depth0.1pt}
\def\Zmath{\vcenter{\hbox{\numbers\rlap{\rlap{Z}\kern
    0.8pt\topfleck}\kern 2.2pt
    \rlap Z\kern 6pt\botfleck\kern 1pt}}}
\def\Qmath{
    \vcenter{\hbox{\upright\rlap{\rlap{Q}\kern3.8pt\stroke}\phantom{Q}}}}
\def\Nmath{\vcenter{\hbox{\upright\rlap{I}\kern 1.7pt N}}}
\def\Cmath{\vcenter{\hbox{\upright\rlap{\rlap{C}\kern
                   3.8pt\stroke}\phantom{C}}}}
\def\Rmath{\vcenter{\hbox{\upright\rlap{I}\kern 1.7pt R}}}
\def\Z{\ifmmode\Zmath\else$\Zmath$\fi}
\def\Q{\ifmmode\Qmath\else$\Qmath$\fi}
\def\N{\ifmmode\Nmath\else$\Nmath$\fi}
\def\C{\ifmmode\Cmath\else$\Cmath$\fi}
\def\R{\ifmmode\Rmath\else$\Rmath$\fi}
\def\barray{\begin{eqnarray}}
\def\earray{\end{eqnarray}}
\def\beq{\begin{equation}}
\def\eeq{\end{equation}}

\def\n{\noindent}

\def\Li{{\rm Li}}

\def\AA{\leavevmode\setbox0=\hbox{h}
\dimen0=\ht0 \advance\dimen0 by-1ex\rlap{\raise.67\dimen0\hbox{\char'27}}A}

\def\red#1{{\color{red}{#1}}}

\def\Li{{\rm Li}}

\makeatletter
\def\iddots{\mathinner{\mkern1mu\raise\p@
\vbox{\kern7\p@\hbox{.}}\mkern2mu
\raise4\p@\hbox{.}\mkern2mu\raise7\p@\hbox{.}\mkern1mu}}
\makeatother

\def\Li{{\rm Li}}

\theoremstyle{plain}

\theoremstyle{remark}

\def\red#1{{\color{red}{#1}}}

\def\Tbar{\bar{T}}

\def\vep{\varepsilon}

\def\cdd{{\rm cdd}}

\def\mass{{\rm mass}}
\def\TTbar{[T\Tbar]}
\def\L{{\rm L}}
\def\R{{\rm R}}

\def\mhat{\hat{m}}

\def\UV{{\rm UV}}
\def\IR{{\rm IR}}

\def\SLL{S^{\L\L}}
 \def\SRR{S^{\R\R}}
 \def\SLR{S^{\L\R}}
 \def\SRL{S^{\R\L}}
  \def\GLL{G^{\L\L}}
 \def\GRR{G^{\R\R}}
 \def\GLR{G^{\L\R}}
 \def\GRL{G^{\R\L}}

\def\Lr{{\rm Lr}}
 
 \def\vepL{\varepsilon^\L}
 \def\vepR{\varepsilon^\R}
 
 \def\khat{\hat{k}}
 
 \def\Khat{{\hat{K}}}

 \def\xhat{\hat{x}}
  
\def\cftUV{{{\rm cft}_{{\rm UV}}}}
\def\cftIR{{{\rm cft}_{{\rm IR}} }}

\def\Inc{\mathcal{I}}

\def\Phirel{\Phi_{{\rm rel}}^{(0)} }
\def\PhirelUV{\Phi_{{\rm rel}}^{\uv} }

\def\mhat{\hat{m}}

\def\Jab{J_{ab}}

\def\sCM{\mathfrak{s}}

\def\jab{{J_{ab}}}

\def\hcox{h^*}

\def\Gcoset{[G]}

\def\uv{{\rm uv}}

\def\cIR{c_{\IR}}
\def\cUV{c_{\UV}}

\def\Paraf{{\rm Pf}}

\def\k{k}

\def\GammaUV{\Gamma_\UV}

\def\Khato{\Khat^{(0)}}

\def\Khatop{\Khat^{(0')}}

\def\Onev{1_v}

\def\cfto{{{\rm cft}_0}}

\begin{document}

\def\sqp{\tfrac{\sqrt{5} +1 }{2}}
\def\sqm{\tfrac{\sqrt{5} -1 }{2}}

\def\sutwo{{\rm su(2)}}
\def\suthree{{\rm su(3)}}
\def\sufour{{\rm su(4)}}
\def\sun{{\rm su(n)}}

\def\check{\red{Changrim Check}}

\title{
On the classification of UV completions of  integrable \\
 $T \bar{T}$ deformations  of CFT}
\author{
 Changrim Ahn\footnote{ahn@ewha.ac.kr} 
 and Andr\'e  LeClair\footnote{andre.leclair@gmail.com}
}
\affiliation{Ewha Womans University, Department of Physics, Seoul 03760, S. Korea\\
Cornell University, Physics Department, Ithaca, NY 14850, USA} 

\begin{abstract}

It is well understood that 2d conformal field theory (CFT) deformed  by an irrelevant $T\Tbar$ perturbation  of dimension $4$ has universal properties.  
In particular,  for the most interesting cases,   the theory develops a singularity in the ultra-violet (UV),   signifying a shortest possible distance,  with  a Hagedorn transition in applications to string theory.     
We show that by adding an infinite number of higher $\TTbar_{s>1}$  irrelevant operators of positive integer scaling dimension $2(s+1)$ with tuned couplings,   this singularity can be resolved and the theory becomes UV complete with a Virasoro central charge $\cUV > \cIR$ consistent with the c-theorem.   
We propose an approach to classifying the possible UV completions of a given CFT perturbed by $\TTbar_s$ that are integrable. 
The main tool utilized is the thermodynamic Bethe ansatz.   
We study this classification for  theories with scalar (diagonal) factorizable S-matrices.    For the Ising model with 
$\cIR = \tfrac{1}{2}$ we find 3 UV completions based on a single massless Majorana fermion description  with 
$\cUV = \tfrac{7}{10}$ and $ \tfrac{3}{2}$,  which both have $\CN=1$ SUSY and were previously known,  and we argue that these are the only solutions to our classification problem based on this spectrum of particles.     We find 
 3 additional ones with a spectrum of 8 massless particles related to the Lie group $E_8$ appropriate to a magnetic perturbation
with $\cUV = \tfrac{21}{22} , \tfrac{15}{2}$ and $\tfrac{31}{2}$.    We argue that it is likely there are more cases for this $E_8$ spectrum.  
We also study simpler cases based on $\suthree$ and $\sufour$ where we can propose complete classifications.    For $\suthree$ the infrared (IR) theory is the 3-state Potts model with $\cIR = \tfrac{4}{5}$ and we find 3 completions with  $\tfrac{4}{5} < \cUV \leq \tfrac{16}{5}$.       For the $\sufour$ case,  which has $3$ particles and $\cIR =1$,  and we find  $11$ UV completions
with $1 < \cUV \leq 5$,   most of which were previously unknown.

\end{abstract}

\maketitle
\tableofcontents

\section{Introduction}

Suppose we are given a 2d conformal field theory (CFT),  referred to as $\cftIR$,  with Virasoro central charge 
$c \equiv  c_\IR$, 
and we  then consider irrelevant perturbations of this CFT: 
\beq
\label{RelPert}
\CS = \CS_{\cftIR} + \sum_{i \geq 1}   \alpha_i \int d^2 x \, \CO_i (x), 
\eeq
where $\CO_i$ are irrelevant operators of scaling dimension $\Gamma_i \geq 2$  in units of mass.    
Since the perturbing operators are irrelevant,   the CFT describes the infrared (IR)  fixed point of the above model,  hence the label $\cftIR$.    
Here $\Gamma_i = 2 \Delta_i$ where $\Delta_i$ is the standard left/right conformal dimension.   
The couplings $\alpha_i$ have dimension $2-\Gamma_i$.     
We assume there is a single overall mass scale $M$,  such that $\alpha_i \propto M^{2-\Gamma_i}$.    
There always exists an infinite number of possible irrelevant operators $\CO_i$ 
in part because we can always differentiate lower dimension operators.     
The standard and essentially correct thinking is that starting with only a few non-zero $\alpha_i$,  perturbation theory generates an infinite number of additional couplings $\alpha_i$,   
the ultraviolet  (UV) limit does not exist,  and predictability is lost.     
This article explores a fundamental question:   can one tune the couplings 
$\alpha_i$ such that the UV limit exists?    
This general problem is fundamental to  attempts to theorize  beyond the Standard Model physics,
since the central issue is to find a complete and finite UV theory, perhaps with quantum gravity,   that leads to the  low energy Standard Model  under 
renormalization group (RG) flow.         
In quantum gravity, ultraviolet completeness  is often referred to as ``asymptotic safety"  \cite{Weinberg}.     There are many interesting issues 
in connection with these questions.   For instance,  the c-theorem \cite{ctheorem} indicates that RG flow to lower energies is 
``irreversible"  in the sense that some UV degrees of freedom are lost in the flow.     This leads to the basic question: 
``Can we reconstruct the UV theory from our  limited data on  the low energy infrared theory    by somehow  reversing the RG flow?".    In general the answer is obviously no,  thus examples where it is possible,  due perhaps to symmetries like supersymmetry,    are 
intrinsically interesting.      This is the problem studied in this paper in a specialized context.    Theories of the kind studied here without a UV completion are in a sense analogous to the so-called ``Swampland" \cite{Swampland}.

The above question is very  broadly stated,  and in this paper we will limit its scope by imposing some significant additional structure. 
First we limit ourselves to  integrable theories in $2$ spacetime dimensions.   
As we will see,   this restriction still leads to a rich structure that has not been fully explored.   
There are in general an infinite number of  low dimension irrelevant operators $\CO_i$ to consider depending 
on the choice of $\cftIR$.     
However every CFT has a stress-energy tensor with the conventionally normalized left/right components 
$T(z)$ and $\Tbar (\zbar)$, each of dimension $2$ in units of energy.     
We thus restrict the class of models such that the lowest dimension irrelevant operator is 
$\CO_1 = T \Tbar/\pi^2$,  and define $\alpha = \alpha_1$ of mass (energy)  dimension $-2$.    
We thus define the dimensionless coupling $g = - \alpha M^2$. 
The higher dimensional operators $\{ \CO_{i>1}  \}$ will be denoted $\TTbar_{s>1}$  with 
dimension $2(s+1)$ where $s$ is an odd integer,  which will be  based on the possible integrable perturbations of $\cftIR$ 
as described in \cite{SZ}.          
More specifically, $s+1$ will be associated to the  integer ``spin" of a  local conserved quantity where $T  , \Tbar$  in $\TTbar_s$ have  spin $s+1 $  and $-(s+1)$) respectively;     
this will be reviewed in the next section.

As already stated,  for $s=1$,  $\TTbar_1 = T \Tbar/\pi^2$  is associated to the generic 
energy-momentum conservation.     
If all other $\alpha_{s>1} = 0$,   then it is known that  the ground state energy on a circle of 
circumference $R$ has a universal form \cite{SZ,Tateo1},    
which we now review. 
An important probe of any theory  is the ground state energy $E(R)$ on an infinite cylinder of 
circumference $R$, which was studied in \cite{ZTT,Tateo1,SZ}.    
In thermodynamic language the free energy density is
$\CF (T) =  E(R) / R$,  where $R=1/T$ is the inverse temperature. 
It is standard to express this  quantity in terms of a scaling function $c(MR)$ 
\beq
\label{cR}
E(R) = - \frac{\pi}{6} \, \frac{c(MR)}{R}
\eeq
where $M$ can be identified with a physical energy scale, such as the mass of a particle, or the energy scale of massless particles.  
The UV limit is $r \equiv  MR \to 0$,  whereas the IR corresponds to $r\to \infty$.   
For a conformal theory,  $c(MR)$ is scale invariant, i.e. independent of $MR$,  and  for unitary theories is equal to the Virasoro central charge.  
For non-unitary theories it is shifted $c\to c -12 d_0$ where $d_0$ is the lowest scaling dimension of fields.
The quantity $c(MR)$ can be used to track the RG flow.
It has been shown that 
\beq
\label{chGeneral}
c(MR) = \frac{2 \cIR }{1 + \sqrt{1- \frac{2 \pi h}{3} \cIR}} , ~~~~~h \equiv \frac{g}{(MR)^2}
\eeq
where
 $\cIR $ is a constant $-\infty < \cIR  < \infty$ identified as the IR central charge as $MR \to \infty$.     
 This  result was  obtained in  \cite{SZ,Tateo1} based on the inviscid Burgers equation
 \beq
 \label{Burgers}
 \d_\alpha E + E \d_R E = 0 .
 \eeq
 Indeed, one finds that \eqref{chGeneral} satisfies the above differential equation if one identifies
 \beq
 \label{params}
 h = - \frac{\alpha}{R^2}, ~~~~~\Longrightarrow ~~ g = - \alpha \, M^2.
 \eeq
 For  $\alpha < 0$ and $\cIR >0$,  one sees that the ground state energy develops a square-root singularity in the UV when 
 $R^2 < 2 \pi |\alpha|/3$,   indicating a smallest possible distance.    It is evident that this singularity only exists
for negative $\alpha$ if $\cIR >0$,  
 and this is consistent with the c-theorem,  i.e. $c(MR)$ increases toward the UV until the singularity is reached \cite{AL}.

Depending on the context,   the singularity in the UV may or may not be desirable.   In the string context, 
the shortest possible distance is related to the string scale and thus  a  kind of Hagedorn transition,  
 and there is no conceptual reason to try to add additional irrelevant operators 
 \cite{Dubovsky,Dubovsky2,Verlinde,Cardy,Tateo2,Tateo3}.    On the other hand,  in traditional QFT,   the singularity 
 signifies the usual pathology with irrelevant perturbations.   In this paper we take the latter point of view, and try to cure the singularity.

To be more specific and summarize the models studied here,   we consider a 2d CFT  formally defined by the action:
\beq
\label{CSmassless0}
\CS_\alpha  = \CS_{\cftIR}  + \sum_{s\geq 1}  \alpha_s \, \int d^2 x \,  \TTbar_s  
\eeq
where the irrelevant operators $\TTbar_s$ depend on the $\cftIR$  and are defined in \cite{SZ}, as reviewed in the next section.   
These generalized $T\Tbar$ deformations have been considered recently in a somewhat
different context \cite{Hernandez,Doyon,Camilo,Cordova}; the main distinction from our work is that 
here the focus is on massless flows between CFT’s which was not studied in these works. 
The problem we pose and study  is to classify  the possible 
tuned   $\alpha_s$   that have UV completions, i.e. theories that are non-singular in the UV.   
This means that in the UV the theories are {\it relevant} perturbations  of  a different  CFT we denote as  $\cftUV$ with central  charge 
$c_\UV$:
\beq
\label{CSUV}
\CS_\UV  \approx  \CS_{\cftUV}  + \lambda' \int d^2 x \, \PhirelUV  \, .
\eeq
The parameters and characteristics of a UV complete  model are $c_\IR$,  $c_\UV$, the tuned  parameters $\alpha_s$ and the dimension of the relevant perturbation in the UV that leads to the RG flow toward the  IR to $\cftIR$:
\beq
\label{GammaUV}
\Gamma_{\UV} = {\rm dim} \( \PhirelUV \) \leq 2 .
\eeq

The remainder of this  article is organized as follows.    In Section II we define in detail the models we consider and propose the  classification problem.  
The $\TTbar_s$ perturbations lead to massless particles where the conformal invariance is broken by non-trivial Left-Right scattering. 
The structure of these massless left/right CDD factors are described in Section III.   In Section IV we present the thermodynamic Bethe ansatz (TBA)
equations for this class of models and list some generic UV completions,  which we refer to as ``{\bf minimal, diagonal, saturated}".       
Our approach is applied to UV completions of the Ising model in Section V.    The interesting feature here is that there are two possible 
massless scattering descriptions of the $\cIR = \tfrac{1}{2}$ critical Ising model:  one is based on the energy perturbation with a spectrum consisting of a single Majorana fermion,  the other is based on the magnetic perturbation and has $8$ massless particles based on the  Lie group
$E_8$ \cite{E8Smatrix}.         
For the Majorana spectrum we provide a complete classification, where these cases were previously known \cite{AlyoshaTIM,AL2,AKRZ}. 
For the $E_8$ case we find 3 UV completions which are new,  however we cannot argue that these cases are exhaustive since there are too many particles to explore the full space of possibilities at this stage. 
In any case the UV completions we find have $\tfrac{1}{2} < \cUV \leq \tfrac{31}{2}$.      In Section VI  we study simpler cases based on 
$\suthree$ and $\sufour$ which have only 2 and 3 particles respectively.   In these cases we propose complete classifications assuming the restrictions we itemize in detail.    
For the $\sufour$ case with $\cIR =1$,   we find $11$ possible completions with $1 < \cUV \leq 5$.    
All of these massless flows are consistent with the c-theorem \cite{ctheorem}. 
The fact that we find many UV completions that were not previously known indicates that our bottom up approach is constructive.   
The Appendix summarizes our notation for current algebra CFTs based on a general simply laced Lie group $G$, and   their associated cosets and parafermions.

\section{General definition of  models and a  proposed classification problem}

Assume we are given a conformal field theory  $\cfto$ formally defined by the action $\CS_\cfto$.    
For our purposes we first  need to 
provide a massless scattering description of the CFT as follows.   We assume there exists 
some integrable perturbation of the CFT by a relevant operator $\Phirel$  of dimension ${\rm dim} (\Phirel) \leq 2$ in mass units  described by the action
\beq
\label{CSmassive}
\CS_\lambda = \CS_\cfto  + \lambda \int d^2 x \, \Phirel (x)
\eeq
where the massive parameter is given by $\lambda = [{\rm mass }]^{2 - {\rm dim} ( \Phirel ) }$.   
We also assume that the theory defined by $\CS_\lambda$ has a massive spectrum of a finite number of particles of physical mass $m_a$, $a=1, 2, \ldots, N$ where $N$ is the number of particle species.    
Being integrable, the  theory has a factorizable $S$-matrix.   In this paper we assume 
the scattering is diagonal,  so that the two particle scattering is given by 
a scalar function $S_{ab} (\theta)$ where $\theta$ is the difference of the usual rapidities of the two particles.    
For an overview of such theories in a broader context we refer to the book \cite{Mussardo}.    

Before taking the massless CFT limit $\lambda \to 0$,  since the theory is integrable,  there exists an infinite number of 
 conserved local currents satisfying the continuity equations
\beq
\d_\zbar T_{s+1} = \d_z \Theta_{s-1}, ~~~~~~
\d_z \Tbar_{s+1} = \d_\zbar  \bar{\Theta}_{s-1},
\eeq
where $s$ is a positive integer 
with $s+1$ and $-(s+1)$  the spins of $T_{s+1}$ and $\Tbar_{s+1}$ respectively.  
For $s=1$ these are the components of the universal stress-energy tensor and the conserved charges are left and right components of momentum.   
For higher $s$,  the above currents depend on the model,  in particular the choice of $\Phirel$.  
For instance,  the  spectrum of the integers $\{ s \}$ depends on the model in question.  
Smirnov and Zamolodchikov showed that from these one can construct well defined local operators $\TTbar_s$:
\beq
\label{Xsdef}
\TTbar_s  = T_{s+1} \Tbar_{s+1} - \Theta_{s-1} \bar{\Theta}_{s-1}
\eeq
with scaling dimension $(\mass)^{2(s+1)}$.    
More importantly, perturbation by such operators preserves the integrability \cite{SZ}.       
Thus,  we can consider the theory defined by the action 
\beq
\label{CSgeneral}
\CS_{\lambda,\alpha}  = \CS_\lambda + \sum_{s\geq 1}  \alpha_s \, \int d^2 x \,  \TTbar_s  
\eeq
where $\alpha_s$ are coupling constants of scaling dimension $[\mass]^{-2s}$.
{\it We emphasize once again that the operators $\TTbar_s$ for $s>1$ depend on the choice of $\Phirel$.  }

It is known that the perturbation by the irrelevant operators $\TTbar_s$ simply modifies the original 
S-matrix $S_{ab} (\theta) $ by 
a CDD factor \cite{SZ,CDD} to
\beq
\label{Scdd}
S_{ab} (\theta)\quad\to\quad S^\cdd_{ab} (\theta)S_{ab} (\theta),\quad
S^\cdd_{ab} (\theta) = \exp \( {i \sum_s g_s^{ab} \, \sinh(s \theta)} \)
\eeq
where the dimensionless coefficient $g_s^{ab}$ is given by
\beq
\label{gsab}
g_s^{ab} = -  \alpha_s \,  (m_a m_b)^s  \,  h^{ab}_s 
\eeq
for some dimensionless coefficients $h^{ab}_s$. 
This massive CDD factor satisfies the  usual  unitarity and crossing relations:
\beq
\label{CDDrel}
S^\cdd (\theta) S^\cdd (-\theta) = 1, ~~~~~~S^\cdd(i \pi - \theta) = S^\cdd (\theta)
\eeq
where we have suppressed $a,b$ indices.  In general the crossing relation involves $\bar{a}$ which is the anti 
$a$-particle.

\def\mass{{\rm mass}}

With the above ingredients we can finally define precisely the type of model we are interested in.  
We first need to select an integrable perturbation $\Phirel$  of $\cfto$  which determines a massive spectrum $m_a$ and their S-matrices
$S_{ab}$.   The choice of $\Phirel$ is not unique, since there could exist more than one integrable perturbation of $\cftIR$.   
As stated above the specific  $\TTbar_s$ depend on $\Phirel$.  
Although the model \eqref{CSgeneral} is interesting in its own right,  the behavior is complicated by the competition between the relevant and irrelevant perturbations since
\beq
\label{couplingsmass}
\lambda \sim  [{\mass}]^{2 - {\rm dim} ( \Phirel )},    ~~~~ \alpha_s \sim \inv{ [\mass]^{2s} } .
\eeq
Thus  in the deep IR where $[\mass] \to \infty$, the $\alpha_s \to 0$ and the theory is dominated by the relevant perturbation $\lambda$.
On the other hand, 
in the extreme UV,  $[\mass] \to 0$,    $\lambda \to 0$ and the operators $\TTbar_s$ are well defined and these irrelevant perturbations dominate, i.e they control the UV behavior.    At an intermediate energy scale there is  cross-over behavior.  

Now we consider the massless limit $\lambda \to 0$:
 \beq
\label{CSmassless}
\CS_\alpha  = \CS_\cftIR  + \sum_{s\geq 1}  \alpha_s \, \int d^2 x \,  \TTbar_s  
\eeq
where $\cftIR = \cfto$.   
In other words we simply utilize the existence of $\Phirel$ to specify a particular massless scattering description for 
$\cftIR$, and  then forget about it.  
 Besides the massless scattering description of $\cftIR$,  the  other remnant of $\Phirel$ is  the specific operators $\TTbar_s$,  except for 
$s=1$ which is universal.     Since $\TTbar_s$ are irrelevant operators,  the CFT defined by $\CS_\cftIR$ 
describes the {\it infrared} limit of $\CS_\alpha$.   {\it  This is in contrast to $\CS_\lambda$ where $\CS_\cfto$ actually describes the UV limit,  and it is  very important to keep this in mind to avoid confusion.}  
As we stated above,   we expect that $\CS_\alpha$ describes the UV limit of $\CS_{\lambda, \alpha}$;  this was the point of view taken in \cite{AL}.

In the massless  limit where all $m_a$ vanish, one must distinguish between left ($\L$) and right ($\R$) movers. 
By scaling the rapidity by $\theta\to \theta_R+\Lambda$ for  ($\R$) and 
$\theta\to \theta_L-\Lambda$ for  ($\L$) with $\Lambda\to\infty$ with
$m_a e^{\Lambda}=\mhat_a$ finite,\footnote{Therefore, the mass ratios remain unchanged.}
the energy and momentum $(E,p)$ can still 
be parameterized by a ``rapidity" $\theta$: 
\barray
{\rm right ~ movers:} ~~~~~ E_a &=&  p_a= \tfrac{\mhat_a}{2} e^{\theta_\R} \cr
{\rm left  ~ movers:} ~~~~~ E_a&=& - p_a= \tfrac{\mhat_a}{2} e^{-\theta_\L}  .
\earray

If there are no $\TTbar_s$ deformations, ($\alpha_s=0$), \eqref{CSgeneral} is then described 
by $\L\L$ and $\R\R$ scattering matrices:
\beq
\label{confLL}
\SLL_{ab} (\theta) = \SRR_{ab} (\theta) = S_{ab} (\theta)
\eeq
where here $\theta = \theta_{\L,a} - \theta_{\L,b}$ or $\theta_{\R,a} - \theta_{\R,b}$.  
These S-matrices satisfy the usual identities \eqref{CDDrel}.       
However, the $\L\R$ and $\R\L$ scattering matrices become trivial,
$\SRL_{ab} (\theta) = \SLR_{ab} (\theta) = 1$, 
since
$|\theta| = |\theta_{\L,a} - \theta_{\R,b}|\to\infty$.
This would mean the theory is just the conformally invariant $\CS_\cftIR$ if it were not for $\TTbar_s$. 
This general framework for massless scattering was pioneered in \cite{ZZ}.

We now show that when $\alpha_s \neq 0$,  if the massless limit is taken properly the $\L\R$ and $\R\L$ scattering remains non-trivial,  consistent with the fact that conformal invariance is broken by $\TTbar_s$.   Furthermore, the massless $\L\R$ CDD factors follow from
the appropriate massless limit of the massive CDD factors \eqref{Scdd}.    We argue as follows.   
In the massless limit $m_a\to 0$, the $g_s$ in \eqref {gsab} vanish and should be rescaled in such a way that
\beq
\label{gsscaling}
\hat{g}^{ab}_s=g^{ab}_s e^{2s \Lambda}=-  \alpha_s \,  (\mhat_a \mhat_b)^s  \,  h^{ab}_s 
\eeq
is finite.
Now the CDD factors for $\L\L$ and $\R\R$ become $1$ since  the $g_s$ vanish with finite $\theta$ in 
\eqref{Scdd}. 
However, the CDD factors for $\R\L$ and $\L\R$ become nontrivial  due to \eqref{gsscaling}.
Combined together with \eqref{confLL},  the full S-matrices for \eqref{CSgeneral} in the massless limit $\lambda\to 0$
become
\barray
\label{SmatrixLL}
\SRR_{ab}(\theta)&=&\SLL_{ab}(\theta)=S_{ab}(\theta),\\
\label{SLR}
\SRL_{ab} (\theta ) &=& \exp \( i \sum_{s\geq 1}  \hat{g}^{ab}_s  e^{s\theta} /2 \),\qquad
\SLR_{ab}  (\theta) = \exp \( -i \sum_{s \geq 1}  \hat{g}^{ab}_s e^{-s \theta} /2 \),
\earray
where $\theta = \theta_\R - \theta_\L$ and $\theta_\L - \theta_\R$ in \eqref{SLR}, respectively. 
We mention that the result \eqref{SLR} was proposed in \cite{AL} by arguing that one   needs  to factorize the massive CDD factor as 
\beq
\label{CDDfactor}
S^{\rm cdd}_{ab} (\theta) = S^{\L\R}_{ab} (\theta) \cdot S^{\R\L}_{ab} (\theta) , 
\eeq
in order for the TBA equations to converge,  however the argument presented above is more complete and rigorous.

For the massless CDD factors \eqref{SLR},   the  following  unitarity/crossing  relations \eqref{CDDrel}
are satisfied
\beq
\label{CDDmassless}
\SRL(\theta)  \, \SRL (\theta + i \pi) = \SLR(\theta)  \, \SLR (\theta + i \pi) = 1,~~~
\SLR(\theta) \, \SRL(-\theta) = 1.
\eeq
Note that the above relations are satisfied regardless of whether $\hat{g}_s^{ab}$ is real or not. 
In the next section we will show how these S-matrices reproduce the universal result \eqref{chGeneral} for pure $\TTbar_1$ perturbations using the TBA.

In summary the models considered here are defined by  the action \eqref{CSmassless} and the S-matrices \eqref{SmatrixLL} and \eqref{SLR}. 
The classification proposed in the Introduction thus proceeds in three  steps.   

\bigskip

\noindent Given $\cftIR$:

\medskip
\noindent
(i)  List the integrable perturbations $\Phirel$  and the corresponding spectrum and massless S-matrices $\SLL$  and $\SRR$  for 
$\cftIR$.   For many CFTs,  these are already known.

\medskip
\noindent
(ii)  For each case in (i), classify the possible $\hat{g}^{ab}_s$ that lead to a UV completion.  

\medskip
\noindent
(iii)   We restrict to {\rm rational} values of $\cUV$ since these are more easily interpreted in terms of known CFTs.   
Requiring that $\cUV$ be rational is difficult to implement directly since from the TBA this requires some very non-trivial identities satisfied by the Rogers dilog function,  many of which were previously unknown.   

\bigskip
\n Although additional restrictions may ultimately be warranted,  these are the main ones considered in this paper.

\bigskip

All the diagonal scalar S-matrices considered in this paper are based on the Dynkin diagram for a simply laced Lie group $G$,  and are 
 $\TTbar_s$ perturbations of the coset $[G]_1 \equiv G_1 \otimes G_1 /G_2$  where $G_k$ is the current algebra at level $k$,  i.e.
 a WZW model.  The central charge of $[G]_1$ is $c= 2\,  {\rm rank} \, G/(2+ \hcox)$ where
$\hcox$ is the dual Coxeter number.   See the Appendix for a summary of these various cosets and our notation.     
 The $[G]_1$ coset CFT is a minimal model for $G = \sutwo, \suthree, E_6, E_7$ and $ E_8$,  with central charges
 $c = \cIR =\tfrac{1}{2}, \tfrac{4}{5}, \tfrac{6}{7} , \tfrac{7}{10} ,  ~{\rm and}  ~\tfrac{1}{2}$ respectively 
 \footnote{The minimal unitary models  of CFT have $c= 1- \tfrac{6}{(p+2)(p+3)} \leq 1$, $p=1,2,3,  \ldots $ which we will refer to again below. These minimal models correspond to the coset $[\sutwo]_{k=p}$.}.   
 Interestingly,  for the critical Ising model at $c=\half$ there are {\it two}  distinct perturbations based on  both $\sutwo$ and $E_8$.  
 Thus the Ising case is the most interesting due to this duplicity.   
 For Ising,  the two choices in step (i) depend  on whether $\Phirel$ is the energy operator, which is a mass term for the Majorana fermion description,  
or  a magnetic  perturbation by the spin field.   In the first case the spectrum is just a single massless Majorana fermion.   In the second the spectrum consists of 8 particles whose masses are related to the root system of the Lie algebra $E_8$.   
 The methods below apply  also to the $E_6$ and $E_7$ cases   however we will not work out these cases in detail in this paper,  
 but rather focus on the $\sutwo$ and $E_8$ cases for $\TTbar_s$ deformations of Ising.   We will also present some results for
 $\suthree$ and $\sufour$.

\section{Fundamental Massless CDD factors}

%
%
%
%
While the  CDD factors given in \eqref{SLR} are directly connected to the $\TTbar_s$ by \eqref{gsscaling},
it will be important to express the massless CDD factor  $\SRL$ in eq. \eqref{SLR}  in terms of basic building blocks which can handle  the UV behavior of the TBA in a controlled way.   
 
Consider first only a single particle so that we can ignore the $a,b$ indices in $\SRL_{ab}$.   In this section we are concerned only with
$\SRL (\theta) $ which for simplicity we will mainly refer to simply  as $S$,  except in some fundamental formulas.     
The RL S-matrices in \eqref{SLR} satisfy the relations \eqref{CDDmassless}.   
As usual we assume there is a single mass scale  $\mhat$ in  the problem after rescaling the rapidities
defined by $me^{\Lambda}=\mhat$.        
Minimal solutions to the equations \eqref{CDDmassless} were considered by Al. Zamolodchikov \cite{AlyoshaTIM}:  
\beq
\label{SAlyosha}
S(\sCM) =    \prod_j  \, \frac{ i \mu^{(j)}   - \sCM}{i \mu^{(j)}+\sCM } 
\eeq
where here $\sCM$ is the center of mass energy 
\beq
\label{sCM}
\sCM = (p_R + p_L)^2 = \mhat^2 \,  e^\theta
\eeq
where $\theta = \theta_R - \theta_L $ and $\mu \propto \mhat ^2$.  
When there are multiple species of particles,   in terms of rapidity this basic CDD factor generalizes to 
\beq
\label{Smin}
S _{ab} (\theta) = \prod_{j=1}^\jab  \( \frac{ i\, \mu_{ab}^{(j)} -  \mhat_a \mhat_b \, e^\theta}{i  \, \mu_{ab} ^{(j)} + \mhat_a \mhat_b \, e^\theta} \) 
\eeq
for some positive integer $\Jab$.  
The number of these  basic factors can be any positive integer $\Jab$ in general,   however not all have a finite UV limit;  this is central to our proposed classification problem.    
The dimensionless parameters $\mu_{ab}^{(j)}/(\mhat_a \mhat_b)$ are also arbitrary but their real parts
should be positive such  that 
the CDD factors do not introduce additional poles in the physical strip.
As we will see, these factors can lead to a  well-defined UV behavior,  with a specific and calculable  $\cUV$.

Now we need to find conditions for the two different forms of the CDD factors \eqref{CDDmassless} and \eqref{SLR}  to match.
In order to compare with \eqref{SLR},  it is convenient to consider the following function,  which in any case will be needed for the kernels in the TBA integral equations:
\beq
\label{dS}
- i \d_\theta \log S_{ab} (\theta) = 2 \sum_{s \geq 1, {\rm odd} }  (-1)^{(s-1)/2} \, b_s^{ab} \, 
( \mhat_a \mhat_b )^s \, e^{s \theta} 
\eeq
where
\beq
\label{bab}
b_s^{ab} = \sum_{j=1}^{\jab}  \( \mu_{ab}^{(j)} \)^{-s} \,.
\eeq
Comparing with \eqref{gsscaling} and \eqref{SLR}  one obtains an infinite number of relations
\beq
\label{gbab}
b_s^{ab} ( \mhat_a \mhat_b )^s= \frac{s}{4} \, (-1)^{(s-1)/2} \, \hat{g}_s^{ab}
\eeq
for all odd integers $s$. 
This implies
\beq
\label{mualpha}
b_s^{ab} = \frac{s}{4} \, \alpha_s \, (-1)^{(s+1)/2} \, h_s^{ab} \, .
\eeq

  The above formula \eqref{mualpha}  is important since it relates the lagrangian couplings $\alpha_s$ 
to the S-matrix parameters $\mu_{ab}^{(j)}$.   It is important to note that in order to relate  \eqref{CDDmassless} to 
\eqref{SLR} it is important to first consider $\theta < 0$ in $S^{\R\L}$ and $\theta > 0 $ in $S^{\L\R}$ in order for the phases to converge,
and then to analytically continue before comparing with \eqref{CDDmassless},  as pointed out in \cite{AL2}.   
Also,  the above formula \eqref{mualpha}  corresponds to a strong fine-tuning since the infinite number of $\hat{g}_s^{ab}$ should be given by 
a finite set of parameters $\mu_{ab}^{(j)}$.
If these conditions are not met, the UV theories are not well-defined.

The S-matrix can then be written as 
\beq
\label{ST} 
\SRL_{ab} (\theta) = \prod_{j=1}^\jab \, T_{\beta_{ab}^{(j)} } (\theta)
\eeq
where 
\beq
\label{Tdef}
T_\beta (\theta) = 
- \tanh  \( \half  ( \theta - \beta -  \tfrac{i \pi}{2} )  \)
\eeq
and 
\beq
\label{betaab}
\frac{\mu_{ab}^{(j)}}{\mhat_a \mhat_b} \equiv  e^{\beta_{ab}^{(j)}}.
\eeq

The kernel for the TBA equations below can then be expressed as 
\beq
\label{Scosh}
\GRL_{ab} ( \theta) \equiv -i \d_\theta \log \, \SRL_{ab} (\theta) = \sum_{j=1}^\jab \, \inv{\cosh \( \theta - \beta_{ab}^{(j)} \) }\,.
\eeq
For $\beta^{(j)}$ real,  the kernel is real,  as it should be.   However the kernel can still be real if $\beta$'s come in complex conjugate pairs. 
For a purely imaginary pair $\beta^{(j)} = \pm i \pi \alpha $  ($\alpha>0$), the massless CDD factor
becomes
\beq
\label{Falpha}
T_{i \pi \alpha } (\theta) \, T_{-i \pi \alpha} (\theta) = F_{\alpha + 1/2} (\theta),
\quad{\rm with}\quad
F_\gamma (\theta) = \frac{ \sinh \theta - i \sin \pi \gamma}{\sinh \theta + i  \sin \pi \gamma }\,.
\eeq
Note that $F_\gamma (\theta)$ has the same form as a massive CDD factor satisfying \eqref{CDDrel} .
We will also need the kernel based on  $F_\gamma$:
\beq
\label{kernel1}
-i \d_\theta \log F_\gamma (\theta) = \frac{ 4 \cosh \theta \sin \pi \gamma}{\cosh 2 \theta - \cos 2 \pi \gamma}\,.
\eeq

We  collect integrations of these real kernels here since we will need them below:
\beq
 \label{Tint}
 -i \int_{-\infty}^\infty \frac{d\theta}{2 \pi}  \, \d_\theta \log T_\beta (\theta) = \half  ~~~~~~{\rm for} ~ \beta ~ {\rm real} 
 \eeq
and 
\beq
\label{FInt}
-i \int_{-\infty}^\infty \frac{d\theta}{2 \pi}  \, \d_\theta \log F_\gamma (\theta) = 
\begin{cases} 
1 , ~~~~~\, {\rm if} ~ \gamma>0 \\
-1, ~~~{\rm if} ~ \gamma <0\,.
\end{cases}
\eeq
We will show below that the UV theories,  if they exist,   can be partially  identified using  only   these integration constants  and the
integers  $J_{ab}$.
Since the latter are independent of the parameters $\beta,\gamma$,
the UV theories can be classified without referring to explicit values of the parameters $\mu_{ab}^{(j)}$ 
as long  as the $\alpha_s$'s satisfy the fine-tuning conditions \eqref{gbab}.

 \section{General Thermodynamic  Bethe Ansatz  for $\TTbar_s$ deformations }

For massive theories the TBA was developed in \cite{ZTBA,KM}.    We are here concerned with massless flows,   which is different in some important respects.

\subsection{TBA equations for the ground state energy} 
  
 Assume we are given $\SLL_{ab}  = \SRR_{ab} $ and $\SRL_{ab}$, $\SLR_{ab}$.   From these we define the kernels  in the usual way:
\barray
\GLL_{ab}(\theta)&=&\GRR_{ab}(\theta)=-i \d_\theta \log \SRR_{ab}(\theta),\\
\GRL_{ab}(\theta)&=&-i\d_\theta \log \SRL_{ab}(\theta),\quad
\GLR_{ab}(\theta)=  -i\d_\theta \log \SLR_{ab}(\theta)=\GRL_{ab}(-\theta),
\earray
where we used the last relation in \eqref{CDDmassless}.
The standard derivation of the TBA  gives
\barray
\label{TBAR}
\vepR_a (\theta ) &=& \tfrac{\mhat_aR}{2} e^\theta -  \sum_b\left[ \GRR_{ab} \star L^\R _b(\theta ) + \GRL_{ab}  \star L^\L_{b}(\theta )\right], \\
\label{TBAL}
\vepL_a (\theta ) &=& \tfrac{\mhat_aR}{2} e^{-\theta} -  \sum_b\left[ \GLL_{ab} \star L^\L _b(\theta ) + \GLR_{ab}  \star L^\R_{b}(\theta )\right],
\earray
where we used the  short  hand notations $L_a^{\L,\R} = \log \left( 1+ e^{-\vep_a^{\L,\R}} \right)$, 
appropriate to a fermionic TBA. 
Above $\star$ denotes the convolution:  $(G \star L)(\theta) = \int_{-\infty}^\infty d \theta' G(\theta - \theta') L(\theta')/2 \pi $.   
From these equations, it is obvious that $\vepR_a (\theta )=\vepL_a (-\theta )$.
In terms of the pseudo energies $\vep_a$, one can find the scaling function of
the ground state energy from 
\beq
\label{GroudnEnergy}
E_0(R)=-\sum_a\,\frac{\hat{m}_a}{4\pi}\int_{-\infty}^{\infty}\left[
e^{\theta}L^\R_a+e^{-\theta}L^\L_a\right]d\theta=
-\sum_a\,\frac{\hat{m}_a}{2\pi}\int_{-\infty}^{\infty}
e^{\theta}L^\R_a\,d\theta.
\eeq

As a check of our pure $\TTbar_1$  massless CDD factors, we can confirm the result from Burgers equation explained in the Introduction 
 from these massless TBA equations.   
If all $\alpha_{s>1}= 0$, the CDD factor can not be given by basic factors $T_{\beta}$ or $F_{\gamma}$.
Instead, the kernel $\GRL_{ab}$ can be directly computed from \eqref{SLR}
\beq
\GRL_{ab}(\theta)=\GLR_{ab}(-\theta)=\frac{\hat{g}_1^{ab}}{2}e^{\theta},\qquad
\hat{g}_1^{ab}=-\alpha_1\, \mhat_a \mhat_b\,.
\eeq
Inserting these into \eqref{TBAR} and \eqref{TBAL} and using \eqref{GroudnEnergy}, one can 
obtain  the same TBA system without $\TTbar_1$ but with shifted $R$:

\beq
\label{TBAabs1}
\vepR_a (\theta ) =\tfrac{\hat{m}_a\tilde{R}}{2} e^\theta -  \sum_b \GRR_{ab} \star L^\R _b,\qquad \vepL_a (\theta )=\tfrac{\hat{m}_a\tilde{R}}{2} 
e^{-\theta}-\sum_b \GLL_{ab} \star L^\L _b,
\eeq
with
\beq
\tilde{R}=R-2\pi\alpha_1\,E_0(R).
\eeq
Therefore, the ground state energy with $\TTbar_1$ should be given by
that of the CFT with the shifted $R$,
\beq
E_0(R)=E_{\rm cft}(R-2\pi\alpha_1\,E_0(R))=-\frac{\pi c_{IR}}{6(R-2\pi\alpha_1\,E_0(R))}.
\eeq
Solution of this quadratic equation reproduces the Burgers result \eqref{chGeneral}.

For the IR limit $MR\gg 1$ with general $\TTbar_s$, one can solve the TBA as a  power expansion of 
$t=1/(MR)^2$ which is  consistent at least up to   $t^3$.    At this order, there are new contributions from $\alpha_3$.
This shows that the S-matrices and TBA are consistent with general $\TTbar_s$ perturbations.

 \subsection{Plateaux equations and $\cUV$}
 
%
The TBA equations \eqref{TBAR} and \eqref{TBAL} for $\varepsilon_a^{\L,\R}(\theta)$ can be solved 
only numerically for generic scale $R$.
However, in the IR and UV limits they can be solved analytically  from which one can extract 
information on both $\cIR, \cUV$ and in principle $\GammaUV$,  although for the latter, the  computations are more difficult.    

Consider the IR limit $R\to\infty$ first.
Since the driving terms $\mhat_aRe^{\theta}$ in \eqref{TBAR} get large while the convolution terms remain finite, 
the pseudo energies $\varepsilon_a^{\R}(\theta)$ diverge for most values of $\theta$ except
a domain $\theta\ll -\log \mhat_aR$ where $\mhat_aRe^{\theta}$ becomes very small so that
$\varepsilon_a^{\R}(\theta)$ can be finite.
Similarly, $\varepsilon_a^{\L}(\theta)$  in \eqref{TBAL} can be finite  in the domain 
$\theta\gg \log \mhat_aR$.
There is no common domain of $\theta$ where both $\varepsilon_a^{\R}(\theta)$ and 
$\varepsilon_a^{\L}(\theta)$ are finite.
Now because the kernels  $G_{ab}(\theta-\theta')$ in the convolution are exponentially small as  
$|\theta-\theta'|\gg 1$, the convolution integrals with $L_b^{\L}(\theta')$ in \eqref{TBAR} become
negligible and decoupled in the TBA equations for $\varepsilon_a^{\R}$.
Similarly, the convolution with $L_b^{\R}(\theta')$ can be neglected for $\varepsilon_a^{\L}$ in \eqref{TBAL}.
Then, the resulting TBA describe the IR CFT.

The UV limit $R\to 0$ is more complicated.
The driving terms $\mhat_aRe^{\theta}$ vanish and the pseudo energies 
$\varepsilon_a^{\R}(\theta)$ become finite in the domain of $\theta\gg -|\log m_aR|$. 
Similarly, $\varepsilon_a^{\L}(\theta)$ are finite for $\theta\ll |\log \mhat_aR|$.
Therefore, both pseudo energies $\varepsilon_a^{\R,\L}(\theta)$ are coupled in the TBA equations
for $ -|\log \mhat_aR|\ll\theta\ll |\log \mhat_aR|$.
In addition, the pseudo energies $\varepsilon_a^{\L,\R}(\theta)$ become virtually flat,
namely, $\theta$-independent in the above domain of $\theta$.
This ``plateaux'' behavior occurs for most kernels except for some exceptional cases.
If this is the case, $L_b^{\R,\L}$ can be pulled out of  the convolution integrals, leaving the integrals of
the kernels.
Then, the non-linear integral TBA system is simplified to a set of simple algebraic equations for the  plateaux values of the pseudo energies, $\hat{\vep}^{R,L}_{a}$.
We refer to these as plateaux equations.

Since $\vep^{\L} _{a}(\theta) =\vep^{\R} _{a}(-\theta) $, both have the same plateaux values $\hat{\vep}_{a }$.
First consider the IR limit,  where the convolutions terms with   $\GRL_{ab} $ and $\GLR_{ab} $ do
not contribute.
The IR plateaux equations  are 
\beq
\label{IRplateaux}
\inv{\xhat_a}  = \prod_b \( 1+\inv{\xhat_b}  \)^{k_{ab} }
\eeq
with $\xhat_a = e^{{\hat{\vep}} _{a } }$ and 
\beq
\label{kab} 
k_{ab} = \int_{-\infty}^\infty \frac{d \theta}{2 \pi} \, \GRR_{ab}
= \int_{-\infty}^\infty \frac{d \theta}{2 \pi} \, \GLL_{ab}.
\eeq
If the solutions  to the plateaux equations are $\xhat_a = x_a$, the IR
central charge is then 
\beq
\label{cIR}
\cIR = \frac{6}{\pi^2} \sum_a  \Lr \( \inv{1 + x_a} \) 
\eeq
where $\Lr$ is the Rogers dilogarithm function defined by
\beq
\label{Rogers}
\Lr  (z) = \Li_2 (z)  + \half \log |z| \log (1-z)
\eeq
where $\Li_2 (z) = \sum_{n=1}^\infty z^n/n^2 $ is the usual dilogarithm.

Turning on $\SLR$, the UV plateaux equations become
\beq
\label{plateaux}
\inv{\xhat_a}  = \prod_b \( 1+\inv{\xhat_b}  \)^{k_{ab} + \khat_{ab}},
\eeq
in terms of the $\R\L$ exponents of the integrated kernels 
\beq
\label{khatab} 
 \khat_{ab} = \int_{-\infty}^\infty 
\frac{d \theta}{2 \pi} \, \GRL_{ab} = \int_{-\infty}^\infty \frac{d \theta}{2 \pi} \, \GLR_{ab} \,.
\eeq
Based on the analogy with the Ising case considered in Section V,   which involves $F_\gamma (\theta)$ with $\gamma = \alpha + 1/2$, 
where $\alpha$ was interpreted as a marginal deformation,   we will restrict ourselves to $F_\gamma$ factors with $\gamma > 0$ for the remainder of this paper \footnote{This restriction may in principle be relaxed,   however we have not explored this possibility here.}.         
From \eqref{Tint} and \eqref{FInt},  one then sees that $\khat_{ab}$ must be a positive  integer multiple of $1/2$ :
\beq
\label{kabint}
\khat_{ab} =  \frac{\Jab}{2} \,.
\eeq
It will be convenient to express these as elements of the matrices $K, \hat{K}$:
\beq
\label{Kmatrix}
\{ k_{ab} \} = K, ~~~~~ \{ \khat_{ab} \} = \hat{K}\,.
\eeq
The UV central charge is given by
\beq
\label{cUV}
c_{UV} =  \frac{6}{\pi^2}  \sum_a  \,  \(  2\,  \Lr \( \inv{1+\xhat_a} \)  - \Lr \( \inv{1+x_a} \) \)  = \frac{12}{\pi^2}  \sum_a  \,    \Lr \( \inv{1+\xhat_a} \)  - \cIR\,.
\eeq
From the $c$-theorem \cite{ctheorem2}, $c_{UV}>\cIR$ should hold and indeed 
does in the cases presented below.

\bigskip

There are two generic cases we refer to as ``saturated" and ``diagonal":

\bigskip

\n  {\bf Saturation point. } ~ Since $\hat{\vep}_a$ must be real,  $\xhat_a=e^{\hat{\vep}_a}$ must be real and positive.
The smallest possible value of $\xhat_a$ is zero,  i.e. $\hat{\vep}_a = - \infty$.      
This happens if $\Khat = \Khato$ satisfying 
 \beq
 \label{sat1}
 K + \Khato= 1,  ~~~~~~\Longrightarrow ~~ \inv{\xhat_a}  =1+\inv{\xhat_a}   ~ ~~  \Longrightarrow ~~
 \xhat_a=0 ~ ~\forall a \,.
 \eeq
 This saturated limit gives the largest possible value of $\cUV$:
 \beq
 \label{cUVmax}
{\rm max} \,  \cUV  = 2\ (\# {\rm particles} ) - \cIR = 2\ {\rm rank} \, G - \cIR\,.
 \eeq
In this equation,  for the cases below based on the Dynkin diagram for the group $G$,  
$\# {\rm particles} = {\rm rank} \, G$.  

\bigskip

\n {\bf Diagonal case.} ~~   This is another generic solution.    
If
\beq
\label{diagonal}
\Khat = - K, ~~~~~~\Longrightarrow ~~  \inv{\xhat_a}  =1 ~~ \forall a ,
\eeq
which gives
\beq
\label{cUVdisagonal}
{\rm diagonal ~ case}: ~~~ \cUV =  (\# {\rm particles} ) - \cIR = {\rm rank} \, G - \cIR = c(G_2/G_1)
\eeq
where we have used $12\  \Lr (1/2)/\pi^2 = 1$,  and $G_2/G_1$ is a coset CFT of two WZW models 
based on the same group $G$ but with two levels $1,2$.

\bigskip
\bigskip

The matrix $K$ is completely fixed by the CFT $\cftIR$  with no further freedom.     However 
 the above $\Khato = 1-K$ is not the unique choice of $\Khat$ that leads to the saturated $\cUV$.   
 Any other choice 
 \beq
 \label{KH}
 \Khatop = \Khato + H
 \eeq
 where 
 $\Onev = (1,1,1, \ldots)^T$  is an eigenvector of $H$ with eigenvalue $0$ also has $\xhat_a = 0, ~ \forall a$.   
 Namely 
 \beq
 \label{Hzero}
 H \, \Onev = 0  ~~~~ \iff ~~ \sum_b H_{ab} = 0, ~~~~~~ \Longrightarrow ~~ \xhat_a = 0 ~~ \forall a .
 \eeq
 
 \bigskip

 Let us anticipate some aspects of the cases we will consider below.   All the IR CFT have $\SLL, \SRR$ structure related to 
 an ADE Dynkin diagram.    If $\Inc$ is the incidence matrix of this diagram,  then $K$ is fixed:
 \beq
 \label{KInc}
 K = - \frac{\Inc}{2-\Inc}~~~~ \Longrightarrow ~ \Khato = 1-K = \frac{2}{2-\Inc}\,.
 \eeq
 In general the elements of $\Khato$ are not integer multiples of $1/2$ as we require in \eqref{kabint},   thus we will choose $H$ 
 such that it is.  
 Due to symmetries of plateaux equations,  for the cases we examined in detail,   {\it all} solutions $\xhat_a$ are identical for $\Khato$ and $\Khatop$ with our choice of $H$ since
 \beq
 \label{Hone}
 \prod_b \( 1 + \inv{\xhat_b} \)^{H_{ab}} = 1\,.
 \eeq

\section{UV completions of the Ising model}
 
As already stated, for the Ising model CFT in the IR,  there are two choices in step (i) depending on whether $\Phirel$ is the energy operator which is a mass term for the Majorana fermion description,  
or  a magnetic  perturbation by the spin field.   In the first case the spectrum of the CFT is just a single massless Majorana fermion.   In the second the spectrum consists of 8  massless particles related to the root system of the Lie algebra $E_8$ \cite{E8Smatrix}.    
Based on this,  in this section we will classify some  possible UV completions of the Ising model and will present $6$ of them in some detail,  
however  as we explain, we are unable to claim as yet that these are exhaustive.

\subsection{Energy operator spectrum:  free Majorana fermion}

If  $\Phirel$  is the energy operator,  then since this is just a mass term for the  Majorana fermion the spectrum consists of 
only one type of massless particle,  the massless Majorana fermion itself.  
Thus $\SLL=\SRR =-1$ and $\cIR = \half$.   
The spectrum of spins for the local integrals of motion is known to be 
 $s= ~{\rm odd ~ integer}$.     This is the $\sutwo$ case of the $\sun$ cases studied in section VI.  

Now as explained above, we deform this IR CFT  by the  set of 
$\TTbar_s$  appropriate to the energy perturbation,  but with fine-tuned coefficients $\alpha_s$.  
It remains to specify the possible $\SRL$.     The general form of this S-matrix was presented in Section III,  where here we can ignore the indices $a,b$.     The possibilities for $\SRL$ are constrained by the requirement that the total kernel should be real and its integral  is bounded so that the plateaux equation gives well-defined central charges $\cUV$.     This leads to the condition
\beq
\label{kbound}
\khat=\int_{-\infty}^{\infty}\frac{d\theta}{2\pi}  \,G^{\R\L}(\theta) \le 1\,.
\eeq
This is evident from the ``saturated limit" described in the last section.  
The parameters of the plateaux equation are $\k =0$ and the single parameter $\khat$.    
The results \eqref{Tint} and \eqref{FInt} imply that $\SRL$ can have at most 2 factors in the product $\prod_j$ in \eqref{ST}. 
The complete classification then  has 3 cases:

\bigskip
\n 1. {\bf Minimal case.} ~~  Here there is only one $T_\beta$ factor with $\beta$ real,   
which we can chose to be zero by shifting the rapidity $\theta$
\beq
\label{minimal}
\SRL (\theta) =  T_{0} (\theta),
\eeq
which corresponds to $\khat = 1/2$.   
The solution to the plateaux equation is $\xhat = (\sqrt{5} -1)/2$,  which leads to $\cUV = \tfrac{7}{10}$.     
This is  the flow from the  tricritical Ising to Ising model first obtained by Al. Zamolodchikov \cite{AlyoshaTIM}.  
The $\CN = 1$ supersymmetry in the tricritical Ising model is spontaneously broken and the goldstino 
particle is the Majorana fermion.  
This case is the $[\sutwo]_2$ to $[\sutwo]_1$ coset flow described in the Appendix,  where our notation is explained there.   The present paper describes this flow  for other Lie groups $G$ in much greater detail than previous works,  for instance in  \cite{Ravanini}.       
Based on this,  we know the dimension of  the relevant perturbation in the UV 
is $6/5$ from \eqref{gammaUV}.      Let us summarize this solution: 
\beq
\label{minimal2}
\cIR = \half  ~~~~\cUV = \tfrac{7}{10},  ~~~~ \GammaUV = \tfrac{6}5, ~~~~~\cftUV = [\sutwo]_2 \,.
\eeq
We should mention that this minimal case does not always correspond to lowest possible value of $\cUV$,   as one can see from Table II for the $\sufour$ case.   

\bigskip
\n 2. {\bf Saturated case.} ~~
The bound  \eqref{kbound}  is saturated with two factors:
\beq
\label{saturated}
\SRL (\theta) = \( T_0 (\theta ) \)^2 = F_{\half} (\theta)\,.
\eeq
which corresponds to $\khat =1$.   
The solution to the plateaux equation is $\xhat = 0$.    
This is a somewhat delicate limit since $\Lr (z)$ has a branch cut along $\Re (z) \geq 1$.   Nevertheless,  the limit $\khat \to 1$ can be approached from below,  and  using $12 \, \Lr (1^-)/\pi^2 = 2$ one finds $\cUV = 3/2$.    This CFT also has $\CN =1$ supersymmetry,  and corresponds to the 
current algebra $\sutwo_2$,  i.e. the $\sutwo$  WZW model at level $k=2$.   (Once again see the Appendix for notation). 
This solution was found in \cite{AL2}.

\bigskip
\n  3.  {\bf  Marginally deformed saturated case} ~~
The bound \eqref{kbound} can also be satisfied with the pair of factors $\{ \beta^{(j)} \} =  \{ \pm i \pi \alpha \} $ with $\alpha > 0$, 
which also has $\khat = 1$:
\beq
\label{alphaSat}
\SRL (\theta) = F_{\alpha + \half} (\theta)
\eeq
Thus there is a one parameter deformation of the saturated case which also has $\cUV = 3/2$.   
This case must be an exactly marginal perturbation of the saturated case.   
In fact this  corresponds to a massless $N=1$ super sinh-Gordon model studied in \cite{AKRZ} with
\beq
\alpha=\frac{1-b^2}{2(1+b^2)}
\eeq
where $b$ is the coupling constant of the theory.   Note that the $\alpha=0$  case in \cite{AL2} corresponds to the self-dual coupling $b=1$.
This result can be confirmed by analyzing the TBA in the UV domain where the effective central charge converges to  $c_{\rm UV}=g$ very slowly in a pattern of inverse powers of $\log({\hat m}R)$, 
typically obtained by the reflection amplitudes of Toda-like exponential interactions.
Just as in the flow from the tricritical Ising to Ising model,   
the massless Majorana fermion, the only particle during the whole RG flow, is a Goldstino which arises from
a spontaneous $\CN=1$ supersymmetry breaking. 
In this case we do not specify $\GammaUV$,  which is in principle determined by the TBA,  since it should depend on the parameter $\alpha$.   
For this reason,  below we will only present solutions without the marginal $\alpha$ deformation,  i.e. we will restrict to $F_\gamma (\theta)$ CDD factors with $\gamma >0$,   typically $\gamma =1/2$.     

Finally,  for this free Majorana case,   the generic diagonal solution described in section IV does not lead to any RG flow since
$K = \Khat =0$,  implying $\cUV = \cIR$.     Given the simplicity of the massless spectrum,  the above 3 cases are a complete classification.

\bigskip

\subsection{The magnetic $E_8$ spectrum}

We now consider a spectrum dictated by the perturbation of  the critical Ising  CFT by the spin field $\sigma(x)$ of dimension $1/8$, 
\beq
\label{SIsingSpin}
\CS_\lambda = S_{\rm Ising} + \lambda \int d^2 x  ~\sigma (x) .
\eeq
The spectrum based on this perturbation is known to consist of 8 massive particles related to the root system of the Lie algebra $E_8$  \cite{E8Smatrix}.
The significance of $E_8$ 
 can be understood using the GKO coset construction and the results in \cite{ABL}.     The completely bootstrapped  massive S-matrix is also known \cite{BCDS},  however we will only refer to its integrated kernels $K$.    
For the CFTs in question,    we refer the reader to the Appendix,  with $G=E_8$.  
 The data one needs is ${\rm dim} \, E_8 = 248$,  ${\rm rank}\,  E_8 = 8$ and $h^* = 30$.    The allowed spectrum of spins $s$ is the odd integers not divisible by $3$ or $5$: 
 \beq
 \label{E8spins}
 \{ s \} = 1,7,11,13,17,19,23, 29  ~~~~{\rm mod} ~ 30. 
 \eeq

This case is considerably more complicated than the $\sutwo$ case above which only had one particle with 
$\{k_{ab} \} = K = 0$.    
Below we will present some simpler cases based on $\sun$;    we present these $E_8$ results here for completeness of this section,
however the reader may benefit from understanding the simpler cases of the next section first.  
 
The spectrum of masses and the completely bootstrapped S-matrix $S_{ab}$ are known \cite{E8Smatrix}.   
For instance,  $m_8/m_1 = 8 \cos^2 (\pi/5)\,  \cos(2 \pi /15)$ etc.   
However we will not need all these details in order to study $\TTbar_s$ deformations;  they are implicit in $S^{\R\R}$.       
The TBA from the $S_{ab}$ is 8 coupled non-linear integral equations and quite complicated.
The analysis is greatly simplified by a universal form of the kernel \cite{ZamADE},
\beq
\label{universal}
\left(\delta_{ab}-\frac{1}{2\pi}\varphi_{ab}(\omega)\right)^{-1}=\delta_{ab}-\frac{1}{2\cosh(\omega/h^*)}\Inc_{ab}
\eeq
where $\varphi_{ab}(\omega)$ is the Fourier transform of the kernel $G_{ab}(\theta)$
and $\Inc_{ab}$ is the incidence matrix for the $E_8$ Dynkin diagram.  
With the labeling of nodes in  Fig.1 one has 
\beq
\label{IncE8}
\Inc = \( 
\begin{smallmatrix}
0 & 0 &1 & 0& 0& 0 & 0 & 0 \\
0 & 0 & 0 & 0 & 0 & 1 & 0 &0   \\
1 & 0 & 0 & 0& 1 & 0 & 0 &0  \\
 0 & 0 & 0 & 0 & 0 & 0 & 0 &1 \\
 0 & 0 & 1 & 0 & 0 & 0 & 1 & 0 \\
0 & 1 & 0 & 0 & 0 & 0 & 0 & 1\\ 
0 & 0 & 0 & 0 & 1 & 0 & 0 &1 \\ 
0 & 0 & 0 & 1 & 0 & 1& 1 &0 \\
   \end{smallmatrix} 
   \) \,.
   \eeq

\begin{figure}
\label{E8Dynkin}
\begin{center}
\begin{picture}(320,50)(20,50)
\put(50,50){\circle{10}}
\put(55,50){\line(1,0){35}}
\put(140,55){\line(0,1){35}}
\put(140,95){\circle{10}}
\put(95,50){\circle{10}}
\put(100,50){\line(1,0){35}}
\put(140,50){\circle{10}}
\put(145,50){\line(1,0){35}}
\put(185,50){\circle{10}}
\put(190,50){\line(1,0){35}}
\put(230,50){\circle{10}}
\put(235,50){\line(1,0){35}}
\put(275,50){\circle{10}}
\put(280,50){\line(1,0){35}}
\put(320,50){\circle{10}}
\put(47,35){$2$}
\put(148,91){$4$}
\put(92,35){$6$}
\put(137,35){$8$}
\put(182,35){$7$}
\put(227,35){$5$}
\put(272,35){$3$}
\put(317,35){$1$}
\end{picture}
\end{center}
\caption{$E_8$ Dynkin diagram}
\end{figure}
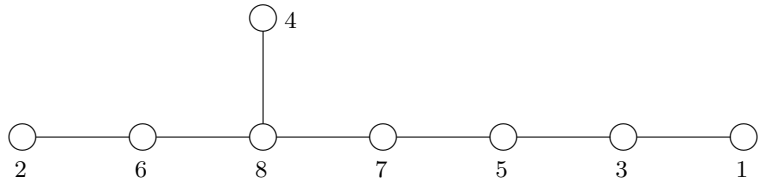

With $\TTbar_s$ deformations, the standard derivation of the universal TBA from \eqref{universal} 
is not valid because $G^{\R\L}_{ab}(\theta)$ need not satisfy similar identities.   
For a complete description 
we should use the original form of the TBA \eqref{TBAR} and  \eqref{TBAL},  
however for some properties this isn't necessary.   
The plateaux equations arising from the TBA in the UV limit need $k_{ab}$ in \eqref{Kmatrix},
which can be computed from $\varphi_{ab}(0)$ 
by taking  $\omega\to 0$ limit for \eqref{universal}
\beq
\label{KE8}
K =  - \frac{\Inc}{2- \Inc} = 
- \(
\begin{smallmatrix}
3 & 4 & 6 &  6 & 8 & 8& 10& 12  \\
4 & 7 &  8 & 10 & 12 &  14, &16 & 20 \\ 
6 & 8 &  11 &  12 &  16 &  16 &  20 &  24 \\ 
6 &  10 &  12 &  15 &  18 &  20 &  24 & 30 \\
8 &  12 &  16 &  18 & 23 &  24 &  30 &  36 \\ 
8 & 14 &  16 &  20 &  24 & 27 &  32 &  40 \\ 
10 &  16 & 20 &  24 &  30 &  32 &  39 &  48 \\ 
12 &  20 &  24 &  30 &  36 &  40 &  48 &  59 \\
 \end{smallmatrix} 
 \) \,.
 \eeq
 The many large integers in the above  $K$ matrix reflects the fact that the S-matrices $\SLL, \SRR$  have many factors analagous to 
 $F_\gamma (\theta)$ factors due to \eqref{FInt}, and most of them give a negative contribution.   

 We limit the possibilities for $\Khat$ based on the saturated limit which has the largest $\cUV$.
 The saturation point is defined as \eqref{sat1}   One solution  is  
\beq
\label{Khat0E8}
\Khato = \frac{2}{2- \Inc} \, .
\eeq  
Since all the entries of $\Khato$ are already half-integers,   there is no need for the matrix $H$ in 
\eqref{KH}.   
 Following insights from the su(2)  case above,  we consider $\Khat$ of the simple form which generates real solutions for $\xhat_a$'s
\beq
\label{Khatn}
\Khat_n = \frac{n}{2} \, \Khato =  n 
\( 
\begin{smallmatrix}
2 & 2 & 3 & 3 & 4 & 4 & 5& 6 \\
2 & 4 & 4 & 5 & 6 & 7 & 8 & 10 \\ 
3 & 4 & 6 & 6 & 8 & 8 & 10 & 12 \\ 
3 & 5 & 6 & 8 & 9 & 10 &  12 & 15 \\
4 &  6 &  8 & 9 &  12 &  12 &  15 &  18 \\ 
4 & 7 & 8 &  10 &  12 & 14 &  16 &  20 \\
5 &  8 &  10 & 12 &  15 & 16 &  20 &  24\\
6 & 10 & 12 &  15 & 18 & 20 & 24 &  30 \\
\end{smallmatrix}
\).
\eeq
From results in Section IV,  in particular \eqref{kabint},  we require entries of $\Khat$ to be integer multiples of $1/2$.   
This allows $n=1/2$,  $n=1$, $n=3/2$  or $n=2$.    Since there are many particles,  we do not present here the detailed 
S-matrices $S^{\R\L}$ in terms of the CDD factors of Section III that lead to the above $\Khat_n$, although we know them;   we will do so below for the simpler case
of $\suthree$.   {\it For  all cases considered in this paper we found such S-matrices and confirmed the plateaux values of $\cUV$ by solving the full TBA equations with the appropriate  rapidity dependent kernels.   The plot in Fig. 3 is typical.  }

Let us write the plateaux equations in a uniform way that depends on $n$, where $n=0$ is the conformal case 
with $\xhat_a = x_a$.    They can be rewritten in the simplest possible way in terms of the sparse matrix $\Inc$.
We found that the following form was most amenable to finding explicit algebraic number solutions:
\beq
\label{plateauxUVIR}
\prod_b \( \xhat_b \)^{\Inc_{ab} - 2 \delta_{ab}} = 
\prod_b \( 1+\inv{\xhat_b} \)^{n \delta_{ab} - \Inc_{ab} }\,.
\eeq
By definition all solutions to these equations are algebraic numbers,   however they don't necessarily have simply presentable  expressions.

\def\s2{{\sqrt{2}}}

\bigskip

\n {\bf Conformal case. }  ~  Here $n=0$ and $\xhat_a = x_a$.     Remarkably the solutions are relatively simple,  only involving the irrational
$\sqrt{2}$
\barray
\label{xcft}
\nonumber
\{ x_1, x_2, \ldots , x_8 \}  &=& 
\{ 2  +  2\s2  , ~5 + 4 \s2 , ~11 + 8 \s2, ~ 16 + 12 \s2 , 
 ~ 42  + 30 \s2, 
  \\  
  \label{AlgebraicE8}
&~& ~~~~~~~~~~ 56 + 40 \s2, ~152 + 108  \s2, 
  ~543 + 384 \s2
  \}\,.
  \earray
  Also remarkably,  $\cUV$ in \eqref{cUV} is exactly equal to $\half$ due to some  evidently  non-trivial identities for the Rogers dilog 
  $\Lr (z)$.     For the cases with non-zero $\Khat$, the dilog identities that are implicit below are certainly unknown.    

\bigskip
\n {\bf Minimal case.} ~
Here $n=1$.    Now the solutions are not as simple as in \eqref{xcft},  but can be reduced to roots  of various quintic polynomials.  
For instance,  $\xhat_1 = 3.228..$ is one root to the polynomial.  
\beq
\label{quintic} 
\xhat ^5 - 2\xhat ^4 -5\xhat ^3 + 2 \xhat ^2 + 4\xhat  +1 = 0.
\eeq
We thus content ourselves with the numerical solution:
\beq
\label{E8minimal}
\{ \xhat_1 , \xhat_2 , \ldots, \xhat_8 \} =
\{ 3.228, ~6.742, ~12.653, ~18.085 , ~39.853, ~51.197, ~ 118.251, \
344.174 \}\,.
\eeq
Again rather remarkably the expression in \eqref{cUV} gives exactly $\cUV = \tfrac{21}{22}$.  
In this case the UV limit is  the $p=9$ unitary minimal model (see the previous footnote).
Referring to the Appendix,   the UV limit is the $[E_8]_2$ coset.  
Let us summarize:
\beq
\label{2122}
\cIR = \half   , ~~ ~~~\cUV = \tfrac{21}{22} , ~~~ \GammaUV = \tfrac{2}{11}, ~~~~~\cftUV = [E_8]_2 
\eeq
Generally,  all the ``minimal" cases considered in this article correspond to the coset $[G]_2$,  and 
have dimension $\GammaUV = 6/(3+\hcox)$ based on \eqref{gammaUV}.

\bigskip

\n {\bf Saturated case.} ~
Here $n=2$ and all $\xhat_a =0$.  
The formula \eqref{cUV} gives $\cUV = \tfrac{31}2$,  which is that of the $E_8$ WZW model at level $k=2$,  here denoted
$E_{8;2}$.
Let us summarize:
\beq
\label{E8saturated}
\cIR = \half  ~~~\cUV = \tfrac{31}{2} , ~~~~~\cftUV = E_{8;2}  
\eeq

Both of the above cases parallel the Majorana case since both UV theories have a fractional supersymmetry with a conserved current of spin 
$31/16$ according to \eqref{Jsusy} and the massless degrees of freedom perhaps can be interpreted as Goldstone particles.  
We will continue to comment on this below.

\bigskip

\n {\bf Diagonal case.} ~
Here $\Khat = -K$, and  this leads to $\xhat_a = 1$  $\forall a$.    As described above and in the Appendix,  the UV CFT can be identified with the 
$\Paraf_2$ parafermion based on $E_8$.   To summarize:
\beq
\label{E8diagonal}
\cIR = \half    ~~~\cUV = \tfrac{15}{2} , ~~~~~\cftUV =   \Paraf_2 \, .
\eeq

\bigskip

A few remarks on $n=1/2, 3/2$ which are  potentially viable.    For $n=1/2$,  there exists solutions 
$\{ \xhat_1, \ldots, \xhat_8 \} = \{ 4.076, \ldots, 665.515 \}$.     However $\cUV = 0.6713$ is irrational to the best of our tests,  thus not a unitary theory. 
Similarly for $n=3/2$.  
As stated in Section II,  we do not include these cases in our  classification.\footnote{We checked rationality up to 30 digits using “Rationalize” in Mathematica. }

Since $\Khat_n$ has entries with large integers,   it is conceivable,  if not likely,   that there are other solutions with 
$\cUV < \tfrac{31}{2}$.    However this is a large space of possibilities to explore,  and we consider it  beyond the scope of this paper.  
     For this reason we cannot claim 
the above $E_8$ cases are complete.   In fact,  in the next section we consider simpler cases with less particles where there are indeed solutions in addition to the generic cases listed in Section IV,  and 
a complete classification is doable under some assumptions.

\section{Some $\sun$ cases}

In this section, we consider $[\sun]_1$ coset CFTs perturbed by $\TTbar_s$.    The case of $\sutwo$ is the Majorana fermion spectrum studied in Section V. 
For $G=\sun$,  ${\rm dim}\, G = n^2 -1$,  ${\rm rank} \, G = n-1$,  and $\hcox = n$.   
The operator $\Phirel$ in Section II is the field $\Phi^{(0)}_{{\rm rel};k=1}$ in \eqref{Phirel0} 
 with dimension $4/(n+2)$.  
 These theories are integrable and described by  known exact diagonal S-matrices for  
$n-1$  massive particles, which can be obtained from those of the $\sun$ affine Toda theories by 
keeping factors independent of the Toda coupling constant
\cite{ABL,BCDS}.    
Following our procedure described in sect.II, we introduce $\TTbar_s$ and corresponding CDD factors. Then,
the massless limit of the S-matrices generate RG flows to UV theories.
Since $\sun$ with small n is far simpler than the $E_8$ case, a detailed classification of  possible UV completions is possible.  

\subsection{Three state Potts: $\suthree$ with $\cIR = 4/5$}

The $[{\rm su(3)}]_1$ coset CFT perturbed by the  relevant operator $\Phirel$  with dimension $4/5$ has
two particles of the same mass. 
The coset CFT has central charge $4/5$.
This CFT is equivalent to $[{\rm su(2)}]_3$ which describes the three state Potts model at  its  critical point.
This theory has $Z_3$ symmetry and is actually the $Z_3$ parafermion defined by another coset 
$\sutwo_3/u(1)$.   The latter parafermion is not the same as the $\suthree$ based parafermion $\Paraf_2$ described in the Appendix.

The matrices  $K$ and $\Khato$  are  given by \eqref{KInc}
\beq
\label{Khato}
K=-\tfrac{1}{3}\left(
\begin{array}{cc}
1&2\\
2&1
\end{array}
\right),   ~~~~~
\Khato = \tfrac{2}{3} \( 
\begin{matrix}
2&1 \\
1&2 \\
\end{matrix}
\).
\eeq
Since $\Khato$ does not consist of only half-integers,   we need a matrix $H$ as described in Section IV: 
\beq
\label{Hmatrix}
H = \tfrac{1}{3} \( 
\begin{matrix}
-1&1 \\
1&-1 \\
\end{matrix}
\),
\eeq
which leads to 
\beq
\label{Khatoprime}
\Khatop= \Khato + H  = \tfrac{1}{2} \( 
\begin{matrix}
2&2 \\
2&2 \\
\end{matrix}
\).
\eeq
With this choice of $H$,  \eqref{Hone} is satisfied since $\xhat_1 = \xhat_2$,  thus
solutions with $\Khato$ and $\Khatop$ are identical.

In this case with only 2 particles,   it is feasible to attempt a complete classification.    Based on $\Khatop$,  we thus  consider 
\beq
\Khat =\tfrac{1}{2}\left(
\begin{array}{cc}
n_1 & n_2 \\
n_2 & n_1
\end{array}
\right).
\eeq
where $n_1, n_2 \in \{ 0,1,2 \}$.    Of course,   $n_1 = n_2 =0$ is just the conformal case with $\cUV = \cIR$.

\bigskip

\n {\bf Minimal case.} ~ Here $(n_1, n_2) = (1,1)$.   
The $S^{\R\L}$ matrices  generate the RG flow from the IR  $[\suthree]_1$  with $\cIR = \tfrac{4}{5}$ to  $[\suthree]_2$  with 
$\cUV = \tfrac{6}{5} $ in the UV.    
We claim that the $S$-matrix elements that generate this flow  are 
\beq
S_{ab}^{\R\L}=T_0(\theta),\quad a,b=1,2 \, .
\eeq
Plateaux values $\xhat_a$ are given in the Table I.   
As it turns out the generic diagonal case described in Section IV also has $\cUV = \tfrac{6}{5}$ and presumably is the same as this minimal case.   Following the Appendix,  the UV CFT can be identified with the parafermion $\Paraf_2$  with integrable deformation by a relevant operator with  $\GammaUV = 1$.

\bigskip

\n {\bf Saturated case.} ~Here $(n_1, n_2) = (2,2)$.   
Now we choose
\beq
S_{ab}^{\R\L}=F_{\tfrac{1}{2}}(\theta),\quad a,b=1,2.
\eeq
Then, in the UV it flows to  a new $\cftUV$  with $c_{\rm UV}=\tfrac{16}5$,  which can be identified as the current algebra  $\suthree_2$.  
As one can see in the $L_a^\R(\theta)$ plot (bell shape) in Fig.2, the naive plateaux assumption is just barely valid in the saturated case. However the numerical solutions of the full TBA confirm that
the plateaux equations determine the  central charge correctly as can be seen in Fig.3.\footnote{One can notice a slower convergence for
the saturated case which reflects the inverse powers of $\log({\hat m}R)$.}  
This means we can use the plateaux equations  even for those cases which do not show robust plateaux behavior.
We also point out that $c(\hat{m} R)$ shown in Fig.3  behaves in a  rather normal fashion. 

\begin{figure}
\begin{center}
\includegraphics[width=350pt, height=250pt]{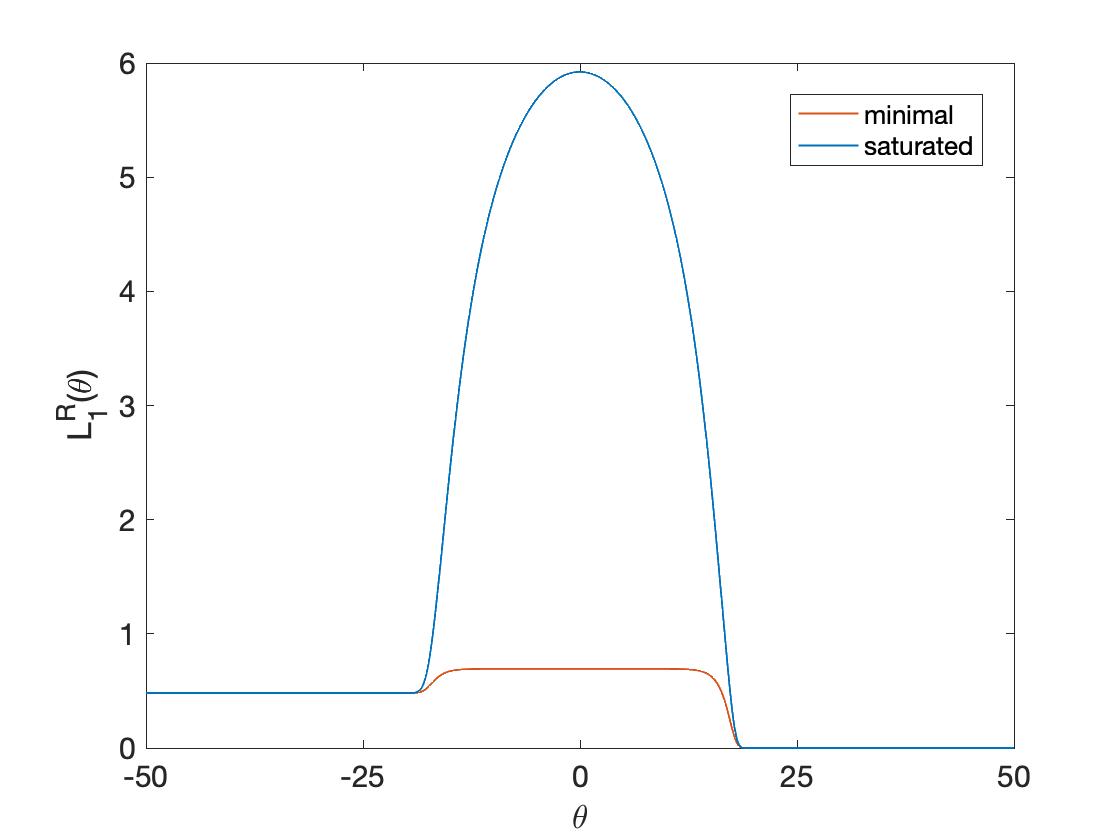}
\end{center} 
\caption{$L_1^{\R}(\theta)$ for minimal and saturated cases of  $\suthree$ at $\mhat R=10^{-7}$. }
\end{figure}

The UV CFT's in the two above cases,  $[\suthree]_2$ and $\suthree_2$ both have a fractional supersymmetry
with a conserved current of spin $8/5$ according to \eqref{Jsusy}.  
In analogy with the $\sutwo$ Ising case,   the massless particles which survive the flow can perhaps  be  interpreted as Goldstone particles for this broken fractional SUSY,   however this suggestion  clearly requires more investigation.

\begin{figure}
\begin{center}
\includegraphics[width=350pt, height=250pt]{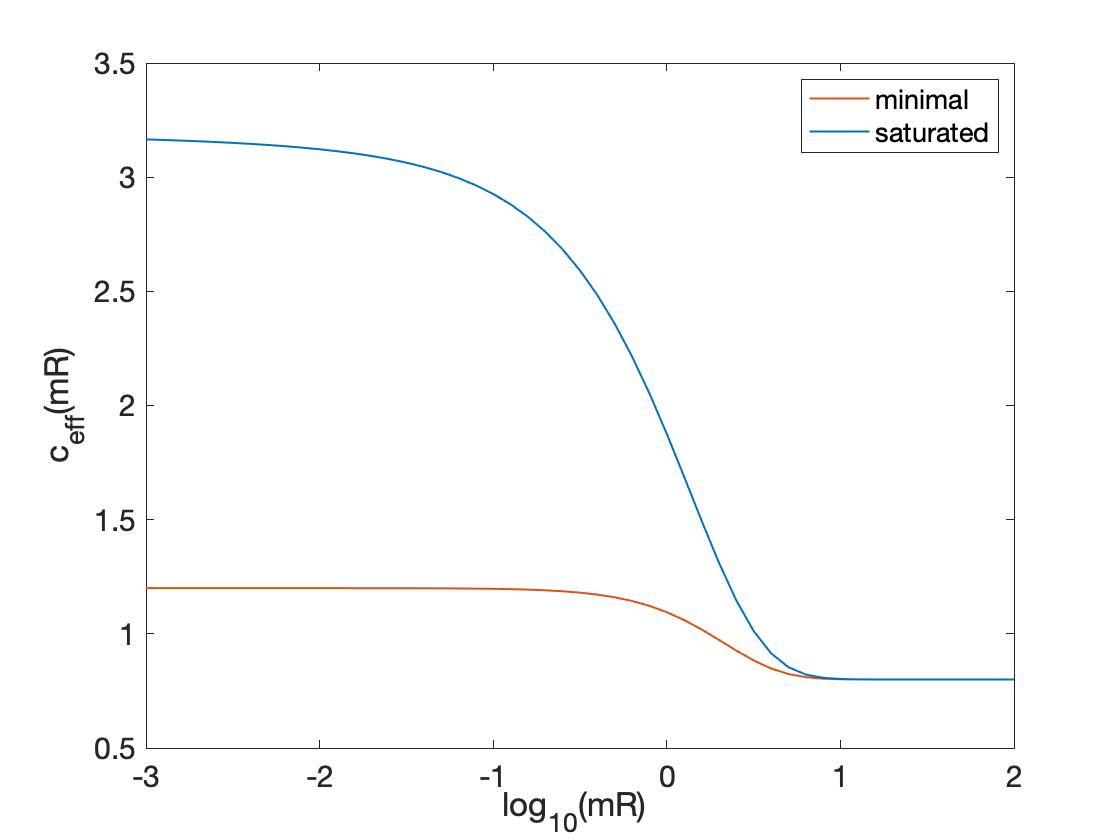}
\end{center} 
\caption{Minimal and saturated RG flows for $\suthree$: $c(\mhat R)$ vs $\log(\mhat R)$ by solving TBA numerically.  The IR central charge
($\mhat R \to \infty$)  is 
$\cIR =4/5$.  One clearly sees that both trajectories arrive to $\cftIR$ from the same direction,  namely  $T\Tbar$. }
\end{figure}

Spanning the space of $\Khatop$,   i.e. $n_1, n_2$,   we find one additional exceptional case with
$\cUV = \tfrac{8}{5}$ where $(n_1, n_2) = (2,1)~ {\rm or} ~(1,2)$. One possible interpretation of this UV CFT is two copies of the IR minimal model each 
with $\cIR = \tfrac{4}{5}$,  however we cannot conclusively make this identification from the value of $\cUV$ alone.   
These three  UV completions,  which we believe to be complete,  are summarized in Table I.   

\begin{table}
\begin{center}
\begin{tabular}{|c|c||c|c|}
\hline\hline
$(n_1, n_2  )$  &$\xhat_1 = \xhat_2$    &   $\cUV $   & $\cftUV$   \\
\hline\hline 
$(0,0)$  & $\sqp$   & $\cUV=\cIR =  \tfrac{4}{5} $ & $\cftUV= \cftIR = [\suthree]_1 $ \\
$(1,1)$  & $1$   & $ \tfrac{6}{5} $ &  $\Paraf_2 $ ~  {\rm ``goldstino" } ~ ({\rm minimal=diagonal})   \\
$(2,1), (1,2)$  & $\sqm$   & $ \tfrac{8}{5} $ & $ [\suthree]_1  \otimes [\suthree]_1 $  ? \\
$(2,2)$  & $0$   & $\tfrac{16}{5} $& $ \suthree_2$ ~~$({\rm saturated}$)  \\
\hline\hline
\end{tabular}
\end{center}
\caption{su(3) results }
\label{su3cUV}
\end{table}

\subsection{$\sufour$ with $\cIR = 1$}

In this case the $K$ matrices \eqref{KInc} are 
\beq
\label{KSu4}
K  = - \tfrac{1}{2} 
\( 
\begin{matrix}
1 &2 & 1 \\
2 & 2 & 2 \\
1 & 2 & 1 \\
\end{matrix}
\),  ~~~~~
\Khato = \tfrac{1}{2} 
\( 
\begin{matrix}
3&2 & 1 \\
2 & 4 & 2 \\
1 & 2 & 3 \\
\end{matrix}
\).
\eeq
It turns out to be useful to introduce an $H$ matrix, as described in Section IVB,  to simplify the form of $\Khato$:
\beq
\label{HSu4}
H = \tfrac{1}{2} 
\( 
\begin{matrix}
-1&0 & 1 \\
0 & 0 & 0 \\
1 & 0 & -1 \\
\end{matrix}
\),
\eeq
which leads to 
\beq
\label{Khat0primeSu4}
\Khatop = \Khato + H  = 
\( 
\begin{matrix}
1&1 & 1 \\
1 & 2 & 1 \\
1 & 1 & 1 \\
\end{matrix}
\).
\eeq

Based on the above $\Khatop$,   we perform a complete classification based on the following space of $\Khat$:
\beq
\label{KhatnsSu4}
\Khat  =    \half  \(
\begin{matrix}
n_1 & n_3 & n_3 \\
n_3 & n_2 & n_3 \\
n_3  &n_3 & n_1 \\
\end{matrix} 
\).
\eeq
Spanning the space of $(n_1, n_2, n_3)$ we find $11$ rational UV completions,   which are summarized in Table II. 
The values $\xhat_1 =0.8019$ and $\xhat_1 = 1.2469$ are solutions to the cubic  equations 
\beq
\label{cubic}
x^3 - 2 x^2 - x +1 =0 ,  ~~~~ x^3 + x^2 -2 x -1 =0
\eeq
respectively.

Let us make some remarks on the results in Table II.   For the generic cases, 
``minimal, diagonal, saturated" we can identify the $\cftUV$ based on arguments in Section IV.   
For the minimal case we know that $\GammaUV = 6/7$ (see Appendix).   
Since $\Paraf_2 = \sufour_2/ \sufour_1$,  it is natural to expect that some of the additional cases with $\cUV$ equal to integer multiples of 
$1/5$ are related to 
$\suthree$ subgroups of $\sufour$,  such as $\sufour_2/ \suthree_1$.  We have tentatively indicated such identifications in the Table. 
However not all exceptional solutions can be explained this way,   in particular 
$\cUV = \tfrac{9}{7}, \tfrac{13}7$,  and other solutions which are multiples of $\tfrac{1}{5}$.    Complete identification of the field content of UV theories in addition to their central charge $\cUV$ is beyond the scope of this paper.        

\begin{table}
\begin{center}
\begin{tabular}{||c|c||c|c||}
\hline\hline
$(n_1, n_2 , n_3 )$  &$(\xhat_1 = \xhat_3, ~\xhat_2)$    &   $\cUV $  & $\cftUV $  \\
\hline\hline 
$(0,0,0)$  & $(2,3)$   & $\cUV= \cIR =1 $  & $\cftUV=\cftIR = [\sufour]_1$  \\
\hline\hline
$(2,0,0)$  & $(1.2469, 4.0489)$   & $ \tfrac{9}{7} $ & ?   \\
\hline\hline
$(2,4,0)$  & $( \sqp, \sqp ) $   & $ \tfrac{7}{5} $ & ?  \\
$(0,2,1)$  &  $ " $   & $  " $  &~ \\
\hline\hline
$ ( 1,2,1 )$  & $ (  1.2469, 1.8019 )$   & $ \tfrac{11}{7} $ & $[\sufour]_2$ ~( minimal) \\
\hline\hline
$ ( 0,0,2 )$  & $ (  1, \sqp  )$   & $ \tfrac{9}{5} $ & $\sufour_2 / \suthree_1 $ ~? \\
$ ( 1,4,1 )$  & $ (  \sqp,\sqm )$   & $ " $  & ~  \\
\hline\hline
$ ( 2,2,1 )$  & $ (  0.8019, 2.2469  )$   & $ \tfrac{13}{7} $ & ? \\
\hline\hline
$ ( 0,2,2 )$  & $ (1,1) $   & $ 2 $ &$\Paraf_2 $ ~ (diagonal ) \\
$ ( 2,4,1 )$  & $ ~"$   & $ " $ &~  \\
\hline\hline
$ ( 1,0,2 )$  & $ (  \sqm, \sqp  )$   & $ \tfrac{11}{5} $ &  ? \\
$ ( 0,3,2 )$  & $ (  1,\sqm )$   & $ " $ & ~  \\
\hline\hline
$ ( 1,2,2 )$  & $ (  \sqm , 1  )$   & $ \tfrac{12}{5} $ & ?  \\
\hline\hline
$ ( 1,3,2 )$  & $ ( \sqm, \sqm  )$   & $ \tfrac{13}{5} $ & ? \\
\hline\hline
$ ( 2,2,2)$  & $(  0,1 ) $   & $ 4 $ & $\sufour_2/\sutwo_1 = \sufour_2/u(1) $~?  \\
\hline\hline
$ ( 2,4,2 )$  & $ (  0,0  )$   & $ 5 $ & $\sufour_2$ ~(saturated )\\
\hline\hline 
\end{tabular}
\end{center}
\caption{su(4) results }
\label{su4cUV}
\end{table}

\subsection{Remarks on general su(n+1)}


The coset $[{\rm su} (n+1)]_1$ deformed by the relevant field $\Phirel$ in Section II, i.e. the field $\Phi^{(0)}_{{\rm rel};k=1}$ in \eqref{Phirel0},  
has $n$ particles with well-known masses and S-matrices.  The complete S-matrices can be easily written down \cite{BCDS}.
From these,  one can find the matrix $K$, which should be $K = -\Inc/(2-\Inc)$ as above.

We have found that the $\hat{K}$ which generates the RG flow from the IR CFT $[{\rm su}(n+1)]_1$
to the UV CFT $[{\rm su}(n+1)]_2$, i.e. the minimal case,   is given by the following $n\times n$ matrix
\beq
\hat{K}=\half \left(
\begin{smallmatrix}
1&1&1&1&1&\cdots&1&1&1&1&1\\
1&2&2&2&2&\cdots&2&2&2&2&1\\
1&2&3&3&3&\cdots&3&3&3&2&1\\
1&2&3&4&4&\cdots&4&4&3&2&1\\
1&2&3&4&5&\cdots&5&4&3&2&1\\
\vdots&\vdots&\vdots&\vdots&\vdots&\cdots&\vdots&\vdots&\vdots&\vdots&\vdots\\
1&2&3&4&5&\cdots&5&4&3&2&1\\
1&2&3&4&4&\cdots&4&4&3&2&1\\
1&2&3&3&3&\cdots&3&3&3&2&1\\
1&2&2&2&2&\cdots&2&2&2&2&1\\
1&1&1&1&1&\cdots&1&1&1&1&1
\end{smallmatrix}
\right).
\eeq
%
%
The above $\Khat$ only describes the generic minimal flow.    Based on the $\suthree, \sufour$ cases above,  we expect many more 
solutions with UV completions that are beyond the scope of this paper to attempt to classify.

\section{Conclusions}

We have shown how  UV singularities in CFTs perturbed by the leading  irrelevant operator $\TTbar_1 = T\Tbar/\pi^2 $
can be resolved by including an infinite number of additional higher dimensional irrelevant operators 
$\TTbar_{s>1}$ with tuned couplings $\alpha_s$.    By requiring integrability,  we have argued that the classification of 
the possible UV completions is a well-defined problem,  and we worked out many cases with diagonal S-matrices. 
We found many UV completions that were previously unknown,  indicating that our proposed classification problem is 
feasible and constructive.

The UV completed theories are in principle completely defined by the S-matrices we propose.   
Our main tool is the TBA which can readily identify the central charge $\cUV$ from the plateaux equations,  as well as the conformal dimension
of the relevant operator perturbation in the UV  by carrying out a more detailed analysis of the TBA in
the UV region which we will not perform in this article.
This information will be essential to figure out 
an independent 
quantum field theory description of the UV field content and its field-theoretic relevant perturbations that lead to the RG flows to the IR that are implicit in the TBAs we propose.    
Thus the bottom up approach from a known IR CFT  to a new UV CFT 
do not as yet completely determine the  field content of the UV  QFT, except in some cases,  
hence some of the ``?" in Tables I and II.   This is perhaps the main open question raised by this work.    
In fact one  should entertain the possibility that the $\cftUV$ that we could not as yet completely identify may perhaps  need to be eliminated by additional restrictions not considered here.

\bigskip

This work raises several other questions which  could lead to interesting developments:

\medskip

\n $\bullet$ ~~ We have clearly stated the restrictions we have imposed on our proposed classification problem that lead for instance to Tables I and II.  
Can these restrictions be relaxed or strengthened,  which could  lead to additional or less possible UV completions?    

\medskip
\n $\bullet$ ~~ Are there additional cases based on the  $E_8$ magnetic spectrum of the Ising model  which are beyond the 3 cases we have found,  and can they be completely classified? 

\medskip

\n $\bullet$ ~~ Although the general ideas presented here  extend to non-diagonal theories,  it would be interesting to work out some examples in detail.  

\medskip

\n $\bullet$ ~~ How important is the role of symmetry?     For some completions of the Ising model,   the massless degrees of freedom,  a 
Majorana fermion,   were understood as Goldstone particles for broken supersymmetry.    For some other cases we suggested that a broken fractional supersymmetry is playing a role.    Can a generalized  Goldstone theorem  be developed?     The fact that the maximal $\cUV$ corresponds to the $G_2$ WZW model suggests that this may be possible.  

\medskip

\n $\bullet$ ~~ Are there interesting physical applications involving the Ising model in a vanishingly small magnetic field with non-zero 
$T\Tbar$ perturbations?   If so,  then  results from Section VB may be useful.       

\bigskip

\section{Acknowledgements}

We would like to thank the organizers and participants of the workshop  APCTP focus program “Exact results
on irrelevant deformations of QFTs”  in October 2021 which led to this collaboration.
AL benefitted from discussions with Denis Bernard and Giuseppe Mussardo.   
This work is supported in part by NRF grant (NRF- 2016R1D1A1B02007258) (CA).

\appendix

\section{Current algebra CFTs and their cosets}

Let $G$ denote a simply laced Lie group (ADE-type).    The data we will need are the dimension of $G$,  its rank, and 
its dual coxeter number $\hcox$.    They are related by 
${\rm dim} \, G   = (\hcox +1)\, {\rm rank} \, G$.  
 Let $G_k$ denote the WZW conformal field theory based on $G$ at level $k$,  where $k$ is a positive integer
 \cite{KniZam,GepWit}.
 It has central charge  
 \beq
 \label{cofk}
c(k)= \frac{k \, {\rm dim}G }{k+\hcox} .
\eeq
Then consider the GKO  coset \cite{GKO}
\beq
\label{coset}
\Gcoset _k =  \frac{G_k \otimes G_1 }{G_{k+1}} , 
\eeq
with central charge $c(\Gcoset_k) = c(k) + c(1) - c(k+1)$.   
For this series of cosets,  the following models are integrable \cite{ABL}:
\beq
\label{SE81}
\CS_\lambda = \CS_{\Gcoset_k}  + \lambda \int d^2 x \, \Phi^{(0)}_{{\rm rel}; k}
\eeq
where $\Phi^{(0)}_{{\rm rel};k} $ is the coset field: 
\beq
\label{Phirel0}
\Phi^{(0)}_{{\rm rel};k} = \[ 
\frac{(k;\circ)\otimes (1;\circ)}{(k+1;{\rm Adj} ) } \] \, .
\eeq
Above,  $\circ$ and  ${\rm Adj}$ denote the scalar and adjoint highest weight representations for the $G_k$ current algebra respectively.
This relevant perturbation has dimension:
\beq
\label{gammaUV}
{\rm dim} \( \Phi^{(0)}_{{\rm rel};k} \)
 =\frac{2(k+1)} {k+1 + h^*}.   
 \eeq

There is at least one UV completion we can anticipate based on conjectured massless flows in coset theories. 
For negative sign of $\lambda$,  the spectrum is massive,  and the S-matrices can be obtained from an RSOS restriction of the 
$G$ affine Toda theory \cite{ABL}.    For positive  $\lambda$,  the model is conjectured to be a massless flow from 
$\Gcoset_k $ to $\Gcoset_{k-1}$.\footnote{See for instance \cite{Ravanini}, and references therein.}
 The flow arrives to the $k-1$ theory via the irrelevant operator 
 \beq
 \label{IrrelArrival}
\Phi^{(0)}_{{\rm irrel};k-1} = \[\frac{ (k-1;{\rm Adj})\otimes (1; \circ) }{(k;\circ)} \] 
\eeq  
of dimension $2(1+ h^*/(k-1+h^*))$. 
Now,  it is important to note that for $k=2$, 
 the operator $(1; {\rm Adj})$ {\it does not exist},  thus for this  massless flow the $\Gcoset_2$ theory should 
arrive to the $\Gcoset_1$ coset CFT  via the $\TTbar_{s}$ operators.    Thus in the UV the theory is described by 
\beq
\label{IsingE82}
\CS_{\rm UV} = \CS_{\Gcoset_2} + \lambda \int d^2 x \,  \Phi^{(0)}_{{\rm rel};2}
\eeq
Based on these ingredients,  in this paper we consider the model defined by \eqref{CSmassless},  
where $\cftIR=\Gcoset_1$. One choice of $\alpha_s$ gives a UV completion which behaves as \eqref{IsingE82} in the UV,  which we refer to as the {\bf minimal} case.    
For this level $k=2$,  $\GammaUV = {\rm dim} ( \Phi^{(0)}_{{\rm rel};2} ) = 6/(3 + \hcox)$,  and we indicate this in the body of the paper. 
 Although this coset flow has been proposed already,   the exact S-matrices and TBA have not previously been worked out at the level of detail presented in this paper.

We also wish to point out that the L,R chiral components of the primary field 
$\Phi^{(0)}_{{\rm irrel};k=2}$ are non-local conserved currents $J_{\rm fsusy}$  for a fractional supersymmetry with dimension (and spin) 
\beq
\label{Jsusy}
\Delta \( J_{\rm fsusy} \) = 1 + \frac{\hcox}{2+ \hcox} .
\eeq
This symmetry also exists in the complete series of cosets $G_\ell \otimes G_2 / G_{\ell +2}$ for all $\ell$,  including the WZW model
$G_2$ which arises in the $\ell \to \infty$ limit.  
For $G=\sutwo$,  where $h^* =2$,  this spin $3/2$ current generates an $\CN=1$ supersymmetry.

In sorting out the $UV$ CFT's it is useful to introduce a free  field content for the current algebra $G_k$.   The cosets can then be described by the introduction of background charges for the bosons.    One needs ${\rm rank} \,G$ bosons and some additional non-abelian parafermions 
based on $G_k$:
\beq
\label{Pfk}
\Paraf_k = G_k / u(1)^{ {\rm rank} G}, ~~~~~~~
c(\Paraf_k) = {\rm rank} G \, \(
\frac{ h^* (k-1)}{k+h^*}  \) .
\eeq
Throughout this paper we do not display the $G$ dependence of $\Paraf_2$ since this evident from the context.

It is interesting to note that 
comparing central charges,  one can potentially make  the identification
\beq
\label{Pfcoset}
\Paraf_2 = G_2/G_1, ~~~c=\frac{ \hcox  {\rm rank} \, G }{2+\hcox} .
\eeq
We have not studied whether the above is an exact equivalence,  nevertheless  we will use this insight for the $\sufour$ cases above with
$G_2/H_1$,  where $H$ is a subgroup of $G$,  to tentatively propose the identification of some UV completions.  
For the $\TTbar_s$ deformations we are here mainly concerned with the $\Paraf_{k=2}$ case.   
Note that for $G=\sutwo$ at level $k=2$,  the  $\Paraf_2$ parafermion is just a Majorana fermion with $c=\half$.

\end{document}